\documentclass{aa} 
\usepackage{graphicx}
\usepackage{txfonts}
\usepackage[usenames]{color} 
\usepackage{tabularx} 
\usepackage{footnote}
\makesavenoteenv{table*}
\usepackage{rotating}
\usepackage{float}
\usepackage{longtable} 
\usepackage{lscape} 

\usepackage{array}
\newcolumntype{H}{>{\setbox0=\hbox\bgroup}c<{\egroup}@{}}
\usepackage{soul}
\usepackage{scalerel}
\usepackage{tikz}
\usetikzlibrary{svg.path}
\definecolor{orcidlogocol}{HTML}{A6CE39}
\tikzset{
  orcidlogo/.pic={
    \fill[orcidlogocol] svg{M256,128c0,70.7-57.3,128-128,128C57.3,256,0,198.7,0,128C0,57.3,57.3,0,128,0C198.7,0,256,57.3,256,128z};
    \fill[white] svg{M86.3,186.2H70.9V79.1h15.4v48.4V186.2z}
                 svg{M108.9,79.1h41.6c39.6,0,57,28.3,57,53.6c0,27.5-21.5,53.6-56.8,53.6h-41.8V79.1z M124.3,172.4h24.5c34.9,0,42.9-26.5,42.9-39.7c0-21.5-13.7-39.7-43.7-39.7h-23.7V172.4z}
                 svg{M88.7,56.8c0,5.5-4.5,10.1-10.1,10.1c-5.6,0-10.1-4.6-10.1-10.1c0-5.6,4.5-10.1,10.1-10.1C84.2,46.7,88.7,51.3,88.7,56.8z};
  }
}

\newcommand\orcid[1]{\href{https://orcid.org/#1}{\mbox{\scalerel*{
\begin{tikzpicture}[yscale=-1,transform shape]
\pic{orcidlogo};
\end{tikzpicture}
}{|}}} \href{#1}{#1}}
\usepackage[breaklinks=true,colorlinks, linkcolor=blue,urlcolor=blue,citecolor=blue]{hyperref}

\usepackage{threeparttablex} 
 
\usepackage[switch]{lineno}

\begin{document}

   \title{Solar energetic electron events measured by MESSENGER and Solar Orbiter}

   \subtitle{Peak intensity and energy spectrum radial dependences: statistical analysis}

   \author{L. Rodríguez-García
          \inst{1}
          \and
          R. Gómez-Herrero\inst{1}
          \and N. Dresing\inst{2}
          \and D. Lario\inst{3}
          \and I. Zouganelis\inst{4}
          \and
          L. A. Balmaceda\inst{3,5} \and \newline A. Kouloumvakos\inst{6}
          \and A. Fedeli\inst{2}
          \and F. Espinosa Lara\inst{1} 
          \and I. Cernuda\inst{1}
          \and G. C. Ho\inst{6}
          \and R. F. Wimmer-Schweingruber\inst{7}
          \and \newline J. Rodríguez-Pacheco\inst{1}
          }
   \institute{Universidad de Alcalá, Space Research Group, Alcalá de Henares, Madrid, Spain \\
              \email{l.rodriguezgarcia@edu.uah.es}
         \and
             Department of Physics and Astronomy, University of Turku, FI-20014 Turku, Finland
             \and
            Heliophysics Science Division, NASA Goddard Space Flight Center, Greenbelt, MD, USA
             \and
             European Space Astronomy Center, European Space Agency, Villanueva de la Cañada, Madrid, Spain
             \and George Mason University, Fairfax, VA, USA
             \and
            The Johns Hopkins University Applied Physics Laboratory, Laurel, MD, USA
             \and
             Institut fuer Experimentelle und Angewandte Physik, University of Kiel, Kiel, Germany
            }

   \date{Received July 20, 2022; accepted November19, 2022}

 
  \abstract
  {We present a list of 61 solar energetic electron (SEE) events measured by the MESSENGER mission and the radial dependences of some parameters associated to these SEE events. The analysis comprises the period from 2010 to 2015, when MESSENGER heliocentric distance varied between 0.31 and 0.47 au. We also show the radial dependencies for a shorter list of 12 SEE events measured in February and March 2022 by spacecraft near 1 au and by Solar Orbiter around its first close perihelion at 0.32 au.  }
   {To study the radial dependences of the electron peak intensity and the energy spectrum of the electron intensity at the time of the SEE event peak intensity taking advantage of multi-spacecraft measurements. } 
   {We compiled the list of SEE events measured by MESSENGER and Solar Orbiter using hourly averages to find the prompt component of the near-relativistic ($\sim$70-110 keV) electron peak intensities, and calculate the peak-intensity energy spectra. We also obtained the peak intensities and energy spectra for the same events as measured by STEREO-A, -B, ACE or Wind spacecraft when one of these spacecraft was in close nominal magnetic connection with MESSENGER or Solar Orbiter to derive the radial dependencies of these SEE parameters.}
   {(1) Due to the elevated background intensity level of the particle instrument on board MESSENGER, the SEE events measured by this mission are necessarily large and intense; most of them accompanied by a CME-driven shock, being widespread in heliolongitude, and displaying relativistic ($\sim$1 MeV) electron intensity enhancements. For this SEE sample we found: (2) The SEE peak intensity shows a radial dependence that can be expressed as R\textsuperscript{$\alpha$}, where the median value of the $\alpha$ index is  $\alpha$\textsubscript{Med}=-3.3$\pm$1.4 for a subsample of 28 events where the nominal magnetic footpoints of the near 0.3 au and 1 au spacecraft were close in heliographic longitude. (3) The mean spectral index $\delta$ of a subset of 42 events where the energy spectrum could be analysed is <$\delta$>=-1.9 $\pm$ 0.3, which is harder than the value found in previous studies using data from spacecraft near 1 au. SEE events observed by Solar Orbiter also display harder energy spectra than prior studies using near 1 au data.}
   {There is a wide variability in the radial dependence of the electron peak intensities, but on average and within uncertainties, the $\propto$ R\textsuperscript{-3} dependence found in previous observational and modelling studies is confirmed. The electron spectral index found in the energy range around $\sim$200 keV ($\delta$200) of the backward-scattered population near 0.3 au measured by MESSENGER is respectively harder by a median factor of $\sim$20\% and $\sim$10\% when comparing to the near 1 au anti-sunward propagating beam and the backward-scattered population.}
   
   \keywords{Sun: particle emission--
                Sun: coronal mass ejections (CMEs) --Sun: flares --
                Sun: corona -- Sun: heliosphere}
   \maketitle
%
\section{Introduction}
\label{sec:Introduc}

Solar energetic electron (SEE) events are sporadic enhancements of electron intensities associated with solar transient activity. At $\sim$1 au, these intensity enhancements are usually observed at near-relativistic energies ($\gtrsim$ 30 keV) and occasionally also at relativistic energies ($\gtrsim$ 0.3 MeV). The mechanisms proposed to explain the origin of solar near-relativistic electron events include acceleration during the processes associated with solar flares \citep{2007Kahler}, magnetic restructuring in the aftermath of coronal mass ejections \citep[CMEs; e.g.][]{2004MaiaPick, 2005Klein}, and/or acceleration at shocks driven by fast CMEs \citep{2002Simnett}.

The passage of interplanetary (IP) shocks at $\sim$1 au is infrequently accompanied by increases in electron intensities \citep{1985TsurutaniLin,2003Lario,2016Dresing}. Therefore, the peak intensity in SEE events is usually observed during the prompt component of the event shortly after its onset. In general, the properties of the SEE events, including the peak intensity measured early in the event, depend on both the processes that accelerate the electrons near the Sun and the time history of the injection of electrons into the IP medium, but also on the transport of these particles from their source up to the spacecraft \citep[e.g.][and references therein]{2009Agueda}.

The observation of SEE events by spacecraft located at heliocentric distances less than 1 au (i.e., closer to the acceleration site) is crucial to understand how solar electrons are injected into IP space. The electron energy spectrum measured near the Sun might resemble the injected spectrum at the flare site or in the CME-shock environment if it is not modified by transport effects. However, transport simulations, which include pitch-angle scattering, show that when injecting particles with a spectrum resembling a single power-law at the Sun, an observer at 0.3 au still observes a single power-law, while at 1 au a broken power-law has formed \citep{Strauss2020}. This is in agreement with the broken power-law spectrum usually found in studies using near 1 au observations \citep[e.g.][]{Dresing2020}, where the change in the spectral shape might be related to stronger IP scattering undergone by higher energy particles ($\gtrsim$ 100 keV). 

The multi-spacecraft observation of SEE events at different heliocentric distances is essential to determine the effects of the particle transport in the properties of the events. In principle, the particle intensities observed by two spacecraft at different helioradii but magnetically connected to the same solar source region depend on (1) how particles are injected onto the IP magnetic field line that connects both spacecraft and (2) how energetic particles are transported from the source region towards the observers \citep{Lario2013}. However, the magnetic connection between two spacecraft cannot always be guaranteed, and the radial dependence of SEE intensities may be different in each individual event. Therefore, statistical analyses over a large number of events are pertinent to characterize the SEE properties, their underlying distributions, possible associations among them and other dependencies, such as for example the radial dependence of the SEE peak intensity or energy spectral index.

In this paper we perform a statistical study of some of the SEE parameters using multi-spacecraft observations of SEE events. In particular, we use energetic electron measurements from 2010 February to 2015 April at different helioradii obtained by the \textit{MErcury Surface Space ENvironment GEochemistry and Ranging} \citep[MESSENGER;][]{Solomon2007MESSENGER} mission near 0.3 au, the twin spacecraft of the \textit{Solar TErrestrial RElations Observatory} \citep[STEREO;][]{Kaiser2008STEREO}, the \textit{Advanced Composition Explorer} \citep[ACE;][]{Stone1998ACE} and the Wind spacecraft \citep{Szabo2015} near 1 au. Combination of these data sets is important because measurements of the radial dependence of electron events are rare, normally due to the scarcity of measurements of SEE events at helioradii < 1 au. Thus, these observations allow us to study some parameters of the SEE events near the Sun and analyse how the IP transport might affect them. In particular, we use MESSENGER and the corresponding spacecraft near 1 au when in close magnetic connection to obtain a radial dependence of the SEE peak intensities and energy spectra of the peak intensity measured in the prompt component of electron events. 

The goals of this study are three. (1) To present all SEE events measured by the MESSENGER mission suitable for analysis (Sect. \ref{sec:SEP_measured_by_MESSENGER}). (2) To determine the radial dependence of the electron peak intensities in the inner heliosphere (Sect. \ref{sec:radial_dependence}). (3) To perform a statistical study of the electron energy spectrum measured at the peak of the event by MESSENGER (Sect. \ref{sec:MESSENGER_peak_spectra}); and to study the radial dependence of the energy spectral indices in the SEE events measured by MESSENGER (Sect. \ref{sec:spectra_comparison}).
The analysis and conclusions from this study are relevant and timely, and can be further developed by ongoing new missions in the inner heliosphere, such as \textit{Solar Orbiter} \citep[][]{Muller2020,Zouganelis2020}, \textit{Parker Solar Probe} \citep[PSP;][]{Fox2016} or \textit{BepiColombo} \citep[][]{2010Benkhoff}. As a preamble, we include in Sect. \ref{sec:solar_orbiter_see_events} SEE events measured by Solar Orbiter near its first close perihelion. During February and most part of March 2022, Solar Orbiter was consistently magnetically connected with STEREO-A along nominal Parker spiral magnetic field lines, so the radial dependence of SEE event properties along the magnetic field can be studied. The rest of the paper is structured as follows. Section \ref{sec:summary_discussion} summarizes and discusses the main findings of the study, and Sect. \ref{sec:Conclusions} outlines our main conclusions. The instrumentation used in this study is introduced in Sect. \ref{sec:Instrumentation}.

\begin{figure*}
\centering
  \resizebox{0.95\hsize}{!}{\includegraphics{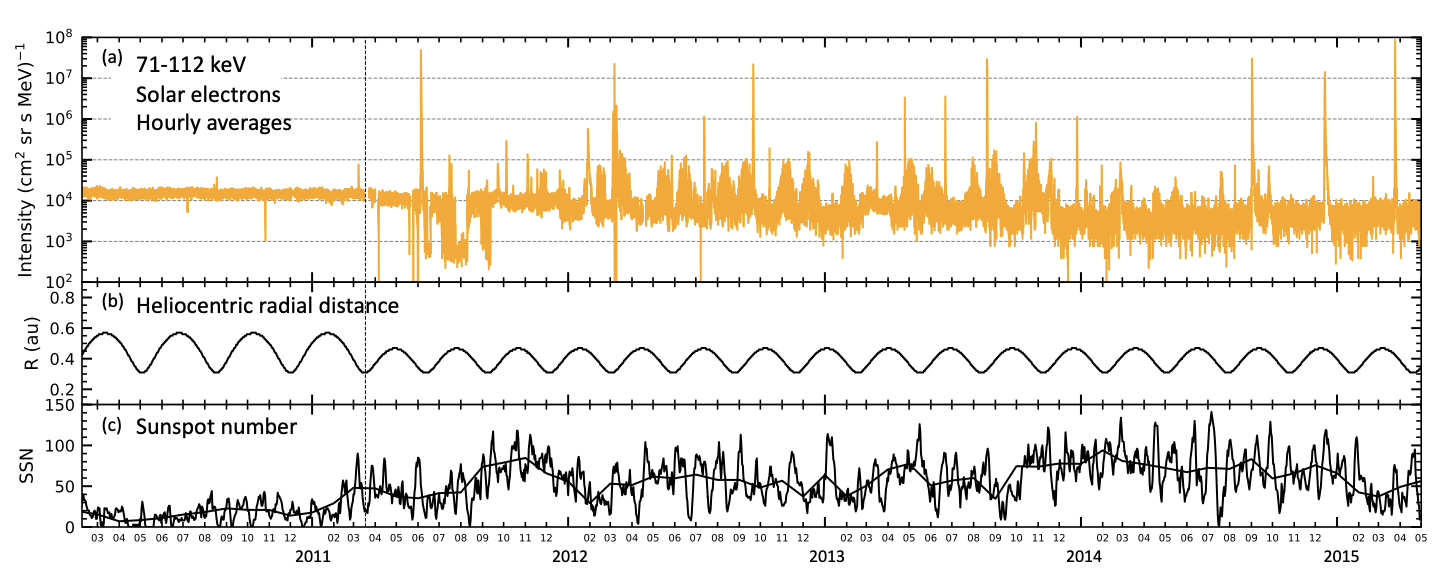}}
     \caption{MESSENGER/EPS data set used for the study. \textit{(a)} Hourly averages of 71-112 keV electron intensities measured by MESSENGER/EPS. The time interval covers from 2010 February 7 to 2015 May 1, where SEE events appear as vertical spikes. The vertical dashed line indicates the time when MESSENGER was injected in the orbit of Mercury. \textit{(b)} Heliocentric distance of the MESSENGER mission during the time of analysis. \textit{(c)} Daily and monthly (smooth line) averages of the sunspot number given by the American Relative Sunspot Number (\url{https://www.ngdc.noaa.gov/stp/space-weather/solar-data/solar-indices/sunspot-numbers/american/lists/}). }
     \label{fig:MESS_mission_71-112keV}
\end{figure*}
\section{Instrumentation} 
\label{sec:Instrumentation}
The statistical study of SEE events from different heliocentric distances requires the analysis of observations from a wide range of instrumentation on board different spacecraft. We used data from MESSENGER, Solar Orbiter, STEREO, ACE, Wind, the \textit{SOlar and Heliographic Observatory} \citep[SOHO;][]{Domingo1995SOHO}, the \textit{Solar Dynamics Observatory} \citep[SDO;][]{Pesnell2012}, and the \textit{Geostationary Operational Environmental Satellites} \citep[GOES;][]{Garcia1994}. 

Remote-sensing observations of CMEs and related solar activity phenomena on the Sun's surface were provided by the Atmospheric Imaging Assembly \citep[AIA;][]{Lemen2012} on board SDO, the C2 and C3 coronagraphs of the Large Angle and Spectrometric COronagraph \citep[LASCO;][]{Brueckner1995} instrument on board SOHO, and the Sun Earth Connection Coronal and Heliospheric Investigation \citep[SECCHI;][]{Howard2008SECCHI} instrument suite on board STEREO. In particular, we used the COR1 and COR2 coronagraphs and the Extreme Ultraviolet Imager \citep[EUVI;][]{Wuelser2004}, part of SECCHI suite. Radio observations were provided by the S/WAVES \citep{Bougeret2008_S/WAVES} investigation on board STEREO and the WAVES \citep{Bougeret1995} experiment on board Wind, and we also consulted the summary plots provided by the Observatoire de Paris-Meudon\footnote{\url{http://secchirh.obspm.fr/}\label{footnote paris-meudon}}. Data from the X-Ray telescopes of the GOES satellites were also used\footnote{\url{https://satdat.ngdc.noaa.gov/sem/goes/data/avg/}\label{goes_data}}. 

In situ energetic particle observations were provided by the Energetic Particle Spectrometer (EPS), part of the Energetic Particle and Plasma Spectrometer \citep[EPPS;][]{Andrews2007EPPS}, on board MESSENGER; the Electron Proton Telescope (EPT) instrument, part of the Energetic Particle Detector \cite[EPD;][]{Rodriguez-Pacheco2020}, on board Solar Orbiter; the Solar Electron and Proton Telescope \citep[SEPT;][]{Mueller2008SEPT} on board STEREO \citep[part of the IMPACT instrument suite,][]{Luhmann2008IMPACT}; the Electron Proton and Alpha Monitor \citep[EPAM;][]{Gold1998EPAM} on board ACE; and the Three-Dimensional Plasma (3DP) and Energetic Particle Investigation \citep[][]{1995Lin3DPWind} on board Wind. Several catalogues were consulted, such as the IP counterpart of CME (hereafter ICME) catalogue at Mercury from the University of New Hampshire\footnote{\url{http://c-swepa.sr.unh.edu/icmecatalogatmercury.html}} \citep{2015Winslow}, the CDAW SOHO LASCO CME catalogue\footnote{\url{https://cdaw.gsfc.nasa.gov/CME_list/}} \citep{Yashiro2004}, the IP shocks catalogue maintained by the University of Helsinki\footnote{\url{http://www.ipshocks.fi/}}  \citep{Kilpua2015}, and the flare list available by the Spectrometer Telescope for Imaging X-rays \citep[STIX on board Solar Orbiter;][]{2020Krucker} data center\footnote{\url{https://datacenter.stix.i4ds.net/view/flares/list}\label{footnote stix}}.

\section{SEE events measured by MESSENGER}
\label{sec:SEP_measured_by_MESSENGER}
In this section we present the SEE events observed by the MESSENGER mission from 2010 February 7 to 2015 April 30 and describe the data source and the criteria for selecting the events.

\subsection{Data source and SEE event selection criteria}
\label{sec:Data_sources_SEP_selection_criteria}

We analysed the MESSENGER/EPS data from 2010 February 7 to 2015 April 30. MESSENGER was initially en route to Mercury on 2004 August 3 and finally inserted into an orbit about the innermost planet on 2011 March 18, until the end of the mission on 2015 April 30. Thus, during the time of analysis, coinciding with the rising, maximum, and declining phase of solar cycle 24, the MESSENGER radial distance varied from 0.31 to 0.47 au. 
 
The EPS instrument measured electrons from $\sim$25 keV to $\sim$1 MeV. The  electron energies chosen here for the SEE events identification and statistical analysis are 71-112 keV. These energies are similar to the 75-105 keV energy band covered by STEREO/SEPT, and the 53-103 keV channel in ACE/EPAM/DE, both used in the analysis of the radial dependence presented in Sect. \ref{sec:radial_dependence}. In addition, the 71-112 keV energy range or similar has been also used in former studies \citep[e.g.][]{Lario2013, 2013Lario_intense, Xie2019}, which facilitates the comparison with previous results. In the case of the analysis of spectra, a portion of the remaining electron energy channels was used, as explained in Sect. \ref{sec:MESSENGER_peak_spectra}. The EPS instrument was mounted on the far-side of the spacecraft with a field of view divided into six sectors pointing in the antisunward direction, so it mostly detected particles moving sunward. Usually, solar energetic particle (SEP) events present a higher particle flux and earlier onset in the sunward-pointing telescope that is aligned with the IP magnetic field \citep[e.g.][]{Gomez-Herrero2021}. Therefore, MESSENGER observations presumably provide a lower limit to the actual peak intensities of the SEP events. In this study, we used data only from sector S02, which is looking above the spacecraft X–Y plane covering 22$^{\circ}$ of the field of view, due to the better signal-to-noise ratio obtained in this sector \citep{2011Hosectors}.  

Figure \ref{fig:MESS_mission_71-112keV}a shows the 71-112 keV electron intensities measured by sector S02 of MESSENGER/EPS from 2010 February 7 to 2015 April 30, where the vertical dashed line indicates the time when MESSENGER was inserted into an orbit about Mercury. Figure \ref{fig:MESS_mission_71-112keV}b shows the variation of the heliocentric distance of MESSENGER during the time of the study, varying from 0.31 to 0.47 au. Figure \ref{fig:MESS_mission_71-112keV}c presents the daily averages of the sunspot number together with its monthly averages (thick black line). In Fig. \ref{fig:MESS_mission_71-112keV}a, we used hourly averages of the particle intensities to improve the statistics of the data, as in  prior studies using MESSENGER data \citep[e.g.][]{Lario2013}. In this compressed time scale, SEE events appear as vertical spikes. The occasional transient bursts of energetic electrons observed in the Mercury's magnetosphere \citep{2011Ho} were excluded in this study. The electronics of the EPS instrument were designed to be able to activate either a large or small pixel for electron detectors, providing a 20-to-1 dynamic range adjustment to maximize the electron detection geometric factor \citep{Andrews2007EPPS}. Because of this adjustment, the instrument background intensity was temporarily reduced in August 2011, as can be seen in Fig. \ref{fig:MESS_mission_71-112keV}a, but it was returned to its original value afterwards. Figures \ref{fig:MESS_mission_71-112keV}a and \ref{fig:MESS_mission_71-112keV}c show that this background level might also be affected by the presence of galactic cosmic rays, showing a gradual decrease as the solar activity increased.

The criteria used to select the SEE events are as follows. (1) The event has to show a clear increase over the background level identified by eye in the 71-112 keV electron channel and (2) a single solar origin should be identified with a distinctive site of the parent solar activity. Regarding the first requirement, and due to the elevated background level of the EPS instrument, the selected events show intensities that are normally above $\sim$10\textsuperscript{4} (cm\textsuperscript{2} sr s MeV)\textsuperscript{-1}. An exception to this is the period of 2011 August, when EPS geometric factor was modified allowing for a transitory detection of less intense events, as discussed above.

\onecolumn
\small
\begin{landscape}
\begin{ThreePartTable}
\begin{TableNotes}

\item \footnotesize{\textbf{Notes. }Columns 1 and 2: Event number and date. Column 3: Type III radio burst onset time. Column 4: Flare location in Stonyhurst coordinates and flare class based on GOES Soft X-ray (SXR) peak flux. Column 5: Longitudinal separation between the flare location and the footpoint of the magnetic field line connecting to MESSENGER, based on a 400 km s$^{\ -1}$ Parker spiral (positive connection angle (CA) denotes a flare source located at the western side of the spacecraft magnetic footpoint). Column 6: MESSENGER radial distance from the Sun. Column 7: 71 -112 keV  electron peak intensity measured by MESSENGER. The pre-event background level is shown in parenthesis. Column 8: Spectral index of peak intensities based on 71 keV to 1 MeV energies. Column 9: 75 to 105 keV electron peak intensity measured by near 1 au spacecraft (STA: STEREO-A; STB: STEREO-B), followed in square brackets by the Corrected peak intensity of a hypothetical 1 au observer with exactly the same CA as MESSENGER. The pre-event background level is given in parenthesis. Column 10: Name of the 1 au spacecraft and the CA difference (CA\textsubscript{near 1au}-CA\textsubscript{MESS}). Column 11: References for the catalogues and studies consulted, as specified in Table \ref{Table:SEPs_References}. * in Col. 1: Widespread SEP event: MESSENGER |CA| or the absolute value of the CA difference with near 1 au spacecraft is $\geq$80$^{\circ}$.  \textsuperscript{++} in Col. 1: The absolute value of the difference between MESSENGER and near 1 au spacecraft CA is: |CA difference| < 20$^{\circ}$. \textsuperscript{+} in Col. 1: The absolute value of the difference between MESSENGER and near 1 au spacecraft CA is: 20$^{\circ}$ $\leq$ |CA difference| $\leq$ 35$^{\circ}$. \textsuperscript{NS} in Col. 2: No CME-driven shock associated to the SEE event. \textsuperscript{ˆ} in Col. 3: Type III radio burst onset time is uncertain due to occultation or multiple radio emission at the same time of the onset of the event. \textsuperscript{§} in Col. 4: The GOES intensity level is deduced from the STEREO/EUVI light curve as explained in \cite{2021Rodriguez-Garcia}. \textsuperscript{$\diamond$} in Col. 8: Presence of 1 MeV electrons. \textsuperscript{\&} in Col. 9: ACE/EPAM/DE 53-103 keV electron intensity divided by an inter-calibration factor of 1.3.

   }

\end{TableNotes}

\setlength{\tabcolsep}{4pt}
\begin{longtable}{ccccHHHccccccc}

\caption{Solar energetic electron events measured by the MESSENGER mission.  }\
\label{Table:SEP_list}\\
\hline
\hline
\multicolumn{4}{c}{Solar event}&Shock&&& \multicolumn{6}{c}{SEE event}&Lists\\
\cline{2-3}  \cline{8-13}
\#&Date &T-III &Flare & speed &width&speed&CA &R& I\textsubscript{max\_MESS} (bg)&$\delta$&I\textsubscript{max\_near\_1au}[I\textsuperscript{'}\textsubscript{corr.}](bg)& s/c: CA diff. \\
& & onset& loc [class]&(3D)&&& MESS & MESS & 71 to 112 keV e&MESS&75 to 105 keV e& \\
& &(UT $\pm$ 5 min)&(deg)&(km s$^{-1}$)&(deg)&(km s$^{-1}$)&(deg)&(au)&(cm\textsuperscript{2} sr s MeV)\textsuperscript{-1}&(-)&(cm\textsuperscript{2} sr s MeV)\textsuperscript{-1}&\\
 \hline
(1)&(2)&(3)&(4)&(5)&(6)&(7)&(5)&(6)&(7)&(8)&(9)&(10)&(11)\\
 \hline
 \endfirsthead
 
\caption{(Continued.)}\\
\hline
\hline
\multicolumn{4}{c}{Solar event}&&&& \multicolumn{6}{c}{SEE event}&Lists\\
\cline{2-3}  \cline{8-13}
\#&Date &T-III &Flare & speed &width&speed&CA &R& I\textsubscript{max\_MESS} (bg)&$\delta$&I\textsubscript{max\_near\_1au}[I\textsuperscript{'}\textsubscript{corr.}](bg)& s/c: CA diff. \\
& & onset& loc [class]&(3D)&&& MESS & MESS & 71 to 112 keV e&MESS&75 to 105 keV e \\
& &(UT $\pm$ 5 min)&(deg)&(km s$^{-1}$)&(deg)&(km s$^{-1}$)&(deg)&(au)&(cm\textsuperscript{2} sr s MeV)\textsuperscript{-1}&(-)&(cm\textsuperscript{2} sr s MeV)\textsuperscript{-1}\\
 \hline
(1)&(2)&(3)&(4)&(5)&(6)&(7)&(5)&(6)&(7)&(8)&(9)&(10)&(11)\\
 \hline
 \endhead
\multicolumn{13}{c}{(Continued on next page.)}

\endfoot

\insertTableNotes  
\endlastfoot

*1\textsuperscript{+} &2010/08/14 &10:00\textsuperscript{ˆ}&N17W052 [C4.4]&960&64&1631&-67 &0.31&2.5$\times$10\textsuperscript{4} (1.6$\times$10\textsuperscript{4})&-& 7.5$\times$10\textsuperscript{1} [1.2$\times$10\textsuperscript{2}] (4.0$\times$10\textsuperscript{1}) &STA: -20$^{\circ}$&a, b, c, f, g \\
*2\textsuperscript{++} &2010/08/18 &05:35&N17W101 [C4.5]&1634&57&1781&-39 &0.31 &3.7$\times$10\textsuperscript{4} (1.5$\times$10\textsuperscript{4}) &-& 3.2$\times$10\textsuperscript{3} [3.1$\times$10\textsuperscript{3}] (3.2$\times$10\textsuperscript{1}) &STA: +1$^{\circ}$&a, b, c, g \\
*3&2011/03/07 &19:55\textsuperscript{ˆ}&N30W048 [M3.7]&2250&51&2505&168  &0.34  & 7.5$\times$10\textsuperscript{4} (1.6$\times$10\textsuperscript{4})  & -1.78$\pm$0.13\textsuperscript{$\diamond$}&-&-&b, c, d, e, g \\
*4\textsuperscript{++} &2011/06/04 &06:50&N16W144 [-]&1086&106&1826& -12 &0.33  &3.1$\times$10\textsuperscript{4} (9.0$\times$10\textsuperscript{3}) &-2.26$\pm$1.14 &7.5$\times$10\textsuperscript{3} [6.8$\times$10\textsuperscript{3}] (7.9$\times$10\textsuperscript{1}) &STA: +3$^{\circ}$&a, b, g \\
*5\textsuperscript{++}&2011/06/04 &21:50\textsuperscript{ˆ}&N16W153 [-]&2200&126&3397&-5  &0.33  & 4.9$\times$10\textsuperscript{7} (2.0$\times$10\textsuperscript{4})  & -1.94$\pm$0.21\textsuperscript{$\diamond$} &6.3$\times$10\textsuperscript{5} [5.9$\times$10\textsuperscript{5}] (3.0$\times$10\textsuperscript{4}) &STA: +5$^{\circ}$&a, b, g \\
*6\textsuperscript{+}&2011/08/02 &06:25\textsuperscript{ˆ}&N15W015 [M1.4]&807&90&1114& 19 &0.46&1.5$\times$10\textsuperscript{3} (2.5$\times$10\textsuperscript{2})&-& 1.0$\times$10\textsuperscript{2} [4.6$\times$10\textsuperscript{2}] (2.3$\times$10\textsuperscript{1}) &STB: +24$^{\circ}$&a, b  \\
*7\textsuperscript{+} &2011/08/04 &03:50&N19W036 [M9.3]&1125&88&2572& 37 &0.46&1.6$\times$10\textsuperscript{3}(5.0$\times$10\textsuperscript{2})  &- & 2.3$\times$10\textsuperscript{2} [7.5$\times$10\textsuperscript{2}] (3.3$\times$10\textsuperscript{1}) &STB: +27$^{\circ}$&a, b, c, d, e, f, g  \\
*8\textsuperscript{++}&2011/09/22 &10:40&N09E089 [X1.4]&1300&81&2206& 90 &0.36&8.1$\times$10\textsuperscript{4} (1.4$\times$10\textsuperscript{4})  &-1.97 $\pm$0.36\textsuperscript{$\diamond$}& 4.5$\times$10\textsuperscript{3} [6.4$\times$10\textsuperscript{3}] (9.4$\times$10\textsuperscript{2}) &STA: +18$^{\circ}$& a, b, c, d, e, f, g   \\
*9&2011/10/04 &12:30\textsuperscript{ˆ}&N26E153 [-]&1358&77&1341& -14 &0.42&2.9$\times$10\textsuperscript{5} (1.0$\times$10\textsuperscript{4})  &-1.88$\pm$0.17\textsuperscript{$\diamond$} &-&-&a, b, c, e, g  \\
10&2011/10/14 &11:00\textsuperscript{ˆ}&N10E140 [-]&889&74&1166& -36 &0.47&2.3$\times$10\textsuperscript{4} (1.2$\times$10\textsuperscript{4})  &-&-&-&a, b\\
*11\ &2011/11/03 &22:15&N09E154 [-]&890&76&1210& -74 &0.44&1.4$\times$10\textsuperscript{5} (9.0$\times$10\textsuperscript{3})  &-1.69$\pm$0.10\textsuperscript{$\diamond$}&-&-&a, b, c, d, e, f, g  \\
12\textsuperscript{+} &2011/11/09 &13:10&N24E035 [M1.1]&1133&45&1446& 34 &0.42&3.6$\times$10\textsuperscript{4} (1.0$\times$10\textsuperscript{4}) &-1.96$\pm$0.28\textsuperscript{$\diamond$} &9.1$\times$10\textsuperscript{2} [9.7$\times$10\textsuperscript{1}] (9.2$\times$10\textsuperscript{1}) &STB: -32$^{\circ}$&a, b, g \\
*13\textsuperscript{++} &2011/11/17 &20:15\textsuperscript{ˆ}&N18E120 [-]&948&106&1254&-71 &0.38&5.8$\times$10\textsuperscript{4} (7.1$\times$10\textsuperscript{3})  &-1.94$\pm$0.26\textsuperscript{$\diamond$}& 7.9$\times$10\textsuperscript{2} [1.0$\times$10\textsuperscript{3}] (2.7$\times$10\textsuperscript{2}) &STB: -11$^{\circ}$&a, g \\
*14\textsuperscript{+}&2012/01/02&14:30&N08W104 [C2.4]&1125&83&1443&-34&0.43&2.1$\times$10\textsuperscript{4} (8.1$\times$10\textsuperscript{3})&-& 6.4$\times$10\textsuperscript{2} [1.7$\times$10\textsuperscript{3}] (3.7$\times$10\textsuperscript{2}) &STA: -29$^{\circ}$& b, g \\
*15\textsuperscript{++}&2012/01/23&03:40&N28W021 [M8.7]&1775&91&2014&-157&0.46&3.4$\times$10\textsuperscript{4} (8.7$\times$10\textsuperscript{3})& -1.78$\pm$0.36\textsuperscript{$\diamond$}&  1.8$\times$10\textsuperscript{4} [1.6$\times$10\textsuperscript{4}] (7.8$\times$10\textsuperscript{1}) &STA: +11$^{\circ}$& b, c, d, e, f, g\\
*16\textsuperscript{++} &2012/01/27 &18:15&N27W078 [X1.7]&1750&70&2468& -108 &0.46&8.7$\times$10\textsuperscript{4} (8.5$\times$10\textsuperscript{3})  &-1.70$\pm$0.19\textsuperscript{$\diamond$}& 1.6$\times$10\textsuperscript{4} [1.1$\times$10\textsuperscript{4}] (1.0$\times$10\textsuperscript{4})&
STA: +19$^{\circ}$&a, b, c, d, e, f, g  \\
*17\textsuperscript{++}&2012/03/04 &11:05&N19E061 [M2.0]&1588&46&1497& -8 &0.31&8.4$\times$10\textsuperscript{4} (8.9$\times$10\textsuperscript{3})  &-2.41$\pm$1.29\textsuperscript{$\diamond$}&  5.5$\times$10\textsuperscript{4} [5.3$\times$10\textsuperscript{4}] (6.2$\times$10\textsuperscript{2}) &STB: +1$^{\circ}$ & b, c, f, g \\
*18\textsuperscript{++} &2012/03/05 &03:35&N17E052 [X1.1]&850&72&2231& -2 &0.31&1.5$\times$10\textsuperscript{6} (4.1$\times$10\textsuperscript{4})  &-1.98$\pm$0.20\textsuperscript{$\diamond$}&7.3$\times$10\textsuperscript{4} [7.5$\times$10\textsuperscript{4}] (2.7$\times$10\textsuperscript{4})&
STB: +4$^{\circ}$&b, d, e, g  \\
*19\textsuperscript{++} &2012/03/07 &00:20&N17E027 [X5.4]&2700&71&3303& 13 &0.31&2.2$\times$10\textsuperscript{7} (1.9$\times$10\textsuperscript{4})  &-2.02$\pm$0.26\textsuperscript{$\diamond$}& 2.1$\times$10\textsuperscript{4} [6.0$\times$10\textsuperscript{4}] (3.7$\times$10\textsuperscript{3}) &STB: +14$^{\circ}$&b, c, d, e, g   \\
*20\textsuperscript{+}&2012/05/17&01:30&N11W076 [M5.1]&1458&75&1807&-76&0.35&8.7$\times$10\textsuperscript{4} (2.0$\times$10\textsuperscript{4})&-& 3.6$\times$10\textsuperscript{2} [5.8$\times$10\textsuperscript{2}] (1.1$\times$10\textsuperscript{1}) &STA: -22$^{\circ}$&b, d, e, f, g\\
*21\textsuperscript{+}&2012/05/26&20:40&N15W121 [-]&1850&55&2665&-75&0.31&1.9$\times$10\textsuperscript{4} (4.0$\times$10\textsuperscript{3})&-1.70$\pm$0.53& 1.3$\times$10\textsuperscript{4} [7.2$\times$10\textsuperscript{3}] (2.3$\times$10\textsuperscript{1}) &STA: +21$^{\circ}$&b, c, f, g\\
*22\textsuperscript{+}&2012/05/27&05:10\textsuperscript{ˆ}&S10E054 [C3.1]&1052&78&958&108&0.31&1.3$\times$10\textsuperscript{5} (2.4$\times$10\textsuperscript{4})&-2.56$\pm$0.96\textsuperscript{$\diamond$}& 1.2$\times$10\textsuperscript{5} [1.7$\times$10\textsuperscript{5}] (8.7$\times$10\textsuperscript{3}) &STA: +23$^{\circ}$&-\\
*23&2012/07/12&15:45\textsuperscript{ˆ}&S15W001 [X1.4]&1393&75&1617& 4 &0.46&1.1$\times$10\textsuperscript{6} (5.5$\times$10\textsuperscript{3})  &-1.95$\pm$0.27\textsuperscript{$\diamond$}&-&-&b, d, f  \\
24&2012/07/17&14:00\textsuperscript{ˆ}&S20W065 [C9.9]&821&50&1245&59&0.46&1.6$\times$10\textsuperscript{4} (2.8$\times$10\textsuperscript{3})&-&-&-&b, f, g \\ 
25&2012/07/19&05:20&S13W088 [M7.7]&1500&71&1897&79&0.46&2.6$\times$10\textsuperscript{4} (7.1$\times$10\textsuperscript{3})&-&-&-&b, g\\
*26&2012/07/23&02:10\textsuperscript{ˆ}&S17W132 [-]&1900&116&2520&116&0.45&5.8$\times$10\textsuperscript{4} (9.5$\times$10\textsuperscript{3})&-1.90$\pm$0.18\textsuperscript{$\diamond$}&-&-&b, d, e, g \\
27&2012/07/28&21:05&S25E055 [M6.1]&792&68&1255&-76&0.44&5.4$\times$10\textsuperscript{4} (4.7$\times$10\textsuperscript{3})&-2.11$\pm$0.42\textsuperscript{$\diamond$}&-&-&-\\
*28 &2012/09/20 &14:55&S15E155 [-]&2600&54&3353& -29 &0.42&2.0$\times$10\textsuperscript{6} (2.5$\times$10\textsuperscript{4})  &-1.91$\pm$0.21\textsuperscript{$\diamond$}&-&-&b, d, e, g  \\
*29\textsuperscript{+}&2012/10/14&00:35&N13E137 [-]&1200&61&1502&-58&0.46&1.9$\times$10\textsuperscript{5} (4.0$\times$10\textsuperscript{3})&-1.93$\pm$0.15\textsuperscript{$\diamond$}
& 6.9$\times$10\textsuperscript{1} [1.3$\times$10\textsuperscript{2}] (4.0$\times$10\textsuperscript{1})
&STB: -25$^{\circ}$&b, g\\
30\textsuperscript{++} &\textsuperscript{NS}2013/03/16 &05:45&
S15W045 [C2.8]&260&61&-& -14 &0.43&2.7$\times$10\textsuperscript{5} (5.0$\times$10\textsuperscript{4}) &-1.92$\pm$0.45\textsuperscript{$\diamond$}&Previous event bg. &ACE:-3$^{\circ}$&-  \\
*31&2013/04/11&07:00&N09E012 [M6.5]&1350&130&1602&-122&0.46&2.2$\times$10\textsuperscript{4} (2.7$\times$10\textsuperscript{3})&-&-&-&b, d, f \\
32 &2013/04/24 &21:40&N10W175 [-]&560&73&1017& 38 &0.40&3.3$\times$10\textsuperscript{6} (7.6$\times$10\textsuperscript{3})  &-2.22$\pm$0.16\textsuperscript{$\diamond$}&-&-&b, d \\
*33\textsuperscript{++} &2013/05/13 &15:55&N11E085 [X2.8]&2000&84&2308& 67 &0.31&2.4$\times$10\textsuperscript{4} (6.3$\times$10\textsuperscript{3})&-1.80$\pm$0.59& 4.1$\times$10\textsuperscript{2} [6.0$\times$10\textsuperscript{2}] (9.2$\times$10\textsuperscript{1}) &STA: +13$^{\circ}$&b, g\\
*34 &2013/06/21 &02:50\textsuperscript{ˆ}&S16E073 [M2.9]&1428&60&2303& -67 &0.46&5.5$\times$10\textsuperscript{5} (4.7$\times$10\textsuperscript{3})  &-1.82$\pm$0.30\textsuperscript{$\diamond$}&-&-& b, f, g \\
35\textsuperscript{+} &\textsuperscript{NS}2013/08/19 &01:20\textsuperscript{ˆ}&N10W162 [-]&-&-&-& -13 &0.32&4.0$\times$10\textsuperscript{4} (1.5$\times$10\textsuperscript{4})&-& No maximum &STA: -29$^{\circ}$&-\\
*36\textsuperscript{+} &2013/08/19 &22:30&N08W178 [M3.3\textsuperscript{§}]&1149&118&1192&-1 &0.32&2.9$\times$10\textsuperscript{7} (1.0$\times$10\textsuperscript{4})  &-1.99$\pm$0.25\textsuperscript{$\diamond$}
& 4.1$\times$10\textsuperscript{4} [8.6$\times$10\textsuperscript{4}] (9.6$\times$10\textsuperscript{2}) &STA: -24$^{\circ}$&b, d, f \\
*37\textsuperscript{+} &2013/10/11 &07:10&N21E103 [M1.5]&875&160&1267&-56 &0.43&1.4$\times$10\textsuperscript{5} (4.6$\times$10\textsuperscript{3})  &-1.92$\pm$0.08\textsuperscript{$\diamond$}&5.9$\times$10\textsuperscript{3} [2.1$\times$10\textsuperscript{3}] (4.4$\times$10\textsuperscript{1})
 &STB: +28$^{\circ}$&b, d, e, g \\
*38&2013/10/25&08:00&S10E073 [X1.7]&500&65&1188&-62&0.36&2.2$\times$10\textsuperscript{5} (1.3$\times$10\textsuperscript{4})&-1.85$\pm$0.16\textsuperscript{$\diamond$}&-&-&b,d, e \\
*39&2013/10/25&15:00&S06E069 [X2.1]&1225&69&1686&-59&0.36&2.8$\times$10\textsuperscript{5} (5.4$\times$10\textsuperscript{4})&-1.89$\pm$0.18\textsuperscript{$\diamond$}&-&-&b, g, f\\
*40 &2013/10/28 &15:10&S08E028 [M4.4]&1400&56&1393&-29 &0.34&8.1$\times$10\textsuperscript{5} (2.1$\times$10\textsuperscript{4})  &-1.97$\pm$0.06\textsuperscript{$\diamond$}&-&-&b, d, e \\
*41&2013/11/19&10:25&S15W069 [X1.0]&1138&52&1361&-41&0.34&6.2$\times$10\textsuperscript{4} (5.4$\times$10\textsuperscript{4})&-1.93$\pm$0.31\textsuperscript{$\diamond$}&-&-&b\\
*42&\textsuperscript{NS}2013/11/30&05:10\textsuperscript{ˆ}&N13W150[-]&-&-&-&2&0.40&1.5$\times$10\textsuperscript{4} (4.9$\times$10\textsuperscript{3})&-
&-&-&b\\
*43&2013/11/30&15:00\textsuperscript{ˆ}&S15E146 [-]&830&48&830&65&0.40&1.6$\times$10\textsuperscript{4} (8.2$\times$10\textsuperscript{3})&-&-&-&b\\
*44\textsuperscript{++}&2013/12/26&03:05&S09E166 [-]&1738&47&1753&-9&0.46&1.1$\times$10\textsuperscript{6} (4.2$\times$10\textsuperscript{3})&-2.02$\pm$0.38\textsuperscript{$\diamond$}& 1.6$\times$10\textsuperscript{4} [2.0$\times$10\textsuperscript{4}] (2.4$\times$10\textsuperscript{1}) &STA: -6$^{\circ}$&b, d, f, g \\
*45\textsuperscript{++}&2014/01/07&18:05&S15W011 [X1.2]&2190&61&2486&145&0.43&3.2$\times$10\textsuperscript{4} (6.1$\times$10\textsuperscript{3})&-& Several events mixed &STA: +16$^{\circ}$&d, e, f, g\\
*46\textsuperscript{++} &\textsuperscript{NS}2014/01/28 &00:30\textsuperscript{ˆ}&S10E081 [C7.6] &-&-&-&-8 & 0.32 &5.9$\times$10\textsuperscript{3} (8.1$\times$10\textsuperscript{2})&-& Ion contamination
 &STB: +17$^{\circ}$&- \\
47\textsuperscript{++} &\textsuperscript{NS}2014/01/28 &05:25\textsuperscript{ˆ}&S14E088 [C9.3] &-&-&-&-16 & 0.32 &2.2$\times$10\textsuperscript{4} (2.7$\times$10\textsuperscript{3})&-2.02$\pm$1.02\textsuperscript{$\diamond$}& Ion contamination &STB:+18$^{\circ}$&- \\
48 &2014/01/30 &16:05 &S13E058 [M6.6] &1450&66&1367&2 & 0.31 &7.4$\times$10\textsuperscript{4} (7.1$\times$10\textsuperscript{3})&-1.82$\pm$0.33\textsuperscript{$\diamond$}&-&-&g\\
49\textsuperscript{+} &2014/02/20 &07:50&S15W073 [M3.0] &1103&70&1328&34 & 0.37 &1.3$\times$10\textsuperscript{4} (1.5$\times$10\textsuperscript{3})&-& 1.3$\times$10\textsuperscript{4} [2.6$\times$10\textsuperscript{3}] (6.7$\times$10\textsuperscript{2})\textsuperscript{\&} &ACE: -22$^{\circ}$&g \\
*50\textsuperscript{++} &2014/02/25 &00:45&S12E082 [X4.9]&2350&69&2431 &-137 &0.40 &5.5$\times$10\textsuperscript{4} (1.2$\times$10\textsuperscript{3})&-1.91$\pm$0.47\textsuperscript{$\diamond$}& 6.7$\times$10\textsuperscript{3} [7.2$\times$10\textsuperscript{3}] (1.2$\times$10\textsuperscript{2})\textsuperscript{\&}
 &ACE: -6$^{\circ}$&d, e, f, g\\
*51\textsuperscript{+} &2014/03/13 &21:40\textsuperscript{ˆ}&N15W140 [-] &498&23&803&81 &0.46 &2.3$\times$10\textsuperscript{4} (3.8$\times$10\textsuperscript{3})&-1.55$\pm$0.31&2.6$\times$10\textsuperscript{2} [5.3$\times$10\textsuperscript{2}] (1.4$\times$10\textsuperscript{2})\textsuperscript{\&}&ACE: +35$^{\circ}$&-\\
52\textsuperscript{+}&2014/08/08&16:15&S10W160 [-]&1035&57&1352&-41&0.33&7.3$\times$10\textsuperscript{4} (6.2$\times$10\textsuperscript{3})&-1.82$\pm$0.21\textsuperscript{$\diamond$}& 1.0$\times$10\textsuperscript{2} [2.1$\times$10\textsuperscript{2}] (4.6$\times$10\textsuperscript{1}) &STA: -23$^{\circ}$&g\\
*53\textsuperscript{++} &2014/09/01 &11:00&N14E127 [-]&1842&77&2947&-44 &0.45&2.9$\times$10\textsuperscript{7} (3.4$\times$10\textsuperscript{3})  &-1.81$\pm$0.03\textsuperscript{$\diamond$}& 4.5$\times$10\textsuperscript{5} [2.5$\times$10\textsuperscript{5}] (1.4$\times$10\textsuperscript{2})
 &STB: +15$^{\circ}$&d, e, g \\
54\textsuperscript{+} &2014/09/05 &06:50&S14E069 [C6.8]&565\textsuperscript{!}&56\textsuperscript{!}&NP& 6 &0.46&8.6$\times$10\textsuperscript{4} (3.9$\times$10\textsuperscript{4})&-2.06$\pm$0.65&No SEE &STB: +23$^{\circ}$&- \\
55\textsuperscript{+} &2014/09/08 &23:55&N12E029 [M4.5]&1120&36&1077& 39 &0.47&2.6$\times$10\textsuperscript{4} (5.4$\times$10\textsuperscript{3})&-& No SEE &STB: +30$^{\circ}$&g \\
*56\textsuperscript{+} &2014/09/10 &17:30&N14E002 [X1.6]&1580&74&1427& 64 &0.47&5.6$\times$10\textsuperscript{4} (1.0$\times$10\textsuperscript{4})&-1.77$\pm$0.16\textsuperscript{$\diamond$}& 9.3$\times$10\textsuperscript{2} [2.2$\times$10\textsuperscript{3}] (3.1$\times$10\textsuperscript{2})
 &STB: +32$^{\circ}$&d, g \\
*57 &2014/09/24 &20:45&N13E179 [-]&1516&76&1651&-139 &0.44&5.3$\times$10\textsuperscript{4} (4.7$\times$10\textsuperscript{3})  &-2.19$\pm$0.13\textsuperscript{$\diamond$}&-&-&d, g \\
58\textsuperscript{++} &2014/12/13 &14:05\textsuperscript{ˆ}&S20W143 [-]&2036\textsuperscript{!}&92\textsuperscript{!}&2519\textsuperscript{!}&-75 &0.46&7.8$\times$10\textsuperscript{6} (3.4$\times$10\textsuperscript{3})  &-1.92$\pm$0.26\textsuperscript{$\diamond$}& No data &STA: -13$^{\circ}$&d, g \\
59\textsuperscript{+}&2015/02/21&09:30\textsuperscript{ˆ}&S40W075 [B4.8]&884\textsuperscript{!}&65\textsuperscript{!}&1088\textsuperscript{!}&-19&0.44&3.8$\times$10\textsuperscript{4} (3.9$\times$10\textsuperscript{3})&-&1.2$\times$10\textsuperscript{3} [2.7$\times$10\textsuperscript{3}] (1.8$\times$10\textsuperscript{2})\textsuperscript{\&} &ACE:+33$^{\circ}$&-\\
60 &2015/03/24 &08:30\textsuperscript{ˆ}&S01W121 [-]&1371\textsuperscript{!}&106\textsuperscript{!}&2102\textsuperscript{!}&-31 &0.43&1.2$\times$10\textsuperscript{6} (1.3$\times$10\textsuperscript{4})  &-1.94$\pm$0.24\textsuperscript{$\diamond$}&-&-&- \\
*61 &2015/04/14 &09:15\textsuperscript{ˆ}&S15W100 [B9] &484\textsuperscript{!}&31\textsuperscript{!}&NP &-119 & 0.32&1.5$\times$10\textsuperscript{4} (4.5$\times$10\textsuperscript{3})&-&-&-&-\\

 \hline

\end{longtable}
\end{ThreePartTable}

\end{landscape}
\twocolumn 

The peak intensity in the prompt component of the event, namely the maximum intensity reached shortly (usually $\lesssim$6 hours) after its onset, is chosen as the maximum intensity. Although electron intensity enhancements associated with the passage of IP shocks are rare \citep{2003Lario,2016Dresing}, by selecting the prompt component of the SEE events, we minimize the possible effect that traveling IP shocks might have on the continuous injection of particles. Therefore, the peak intensity of the SEE event is observed when the respective sources of particles are still close to the Sun. Events showing multiple intensity enhancements are only considered if the first intensity increase reaches a maximum before the second event commences and it can be associated with a single parent solar event.
\subsection{MESSENGER SEE event list}
\label{sec:MESSENGER SEE list}
  \begin{figure}
  \resizebox{0.9\hsize}{!}{\includegraphics{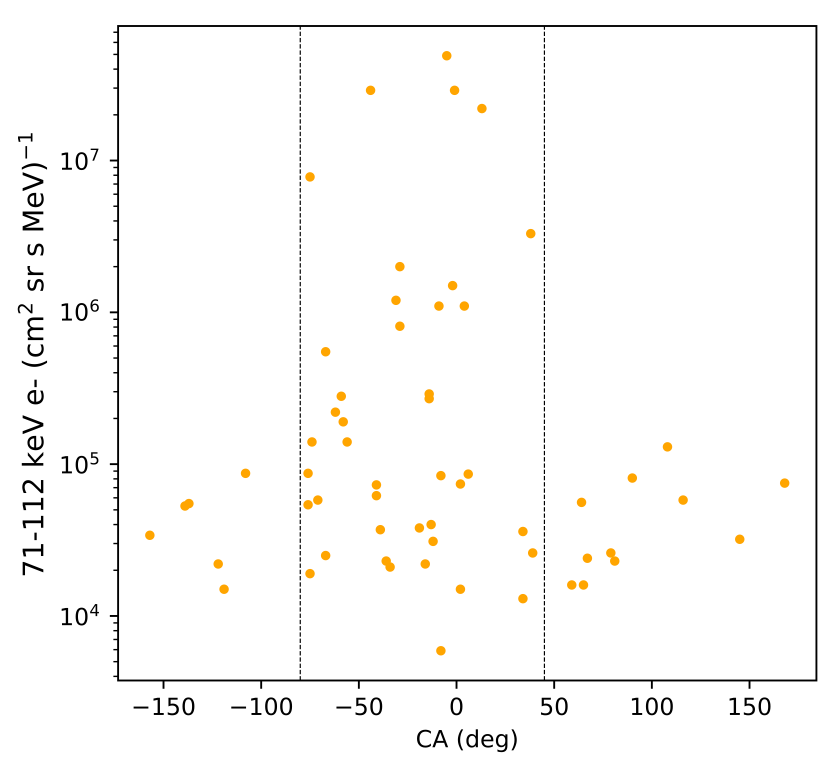}}
     \caption{MESSENGER solar energetic electron peak intensities versus connection angle (CA). The vertical dashed lines indicate the connection angles CA=-80$^{\circ}$ (left) and CA=+45$^{\circ}$ (right).    }
     \label{fig:stats_Int_CA}
\end{figure} 
 Table \ref{Table:SEP_list} shows the list of the 61 SEE events selected in this study. Columns 1-3 identify each SEE event with a number (1), the solar event date (2), and the time of the type III radio burst onset (3), which is determined using the plots available at the Observatoire de Paris-Meudon website\textsuperscript{\ref{footnote paris-meudon}}. We use the symbol (ˆ) to indicate when the type III burst onset time is uncertain due to occultation or multiple radio emission at the same time of the onset of the event. Column (4) provides the location of the solar flare either identified in this study using extreme ultraviolet (EUV) data from STEREO/EUVI and SDO/AIA, or consulted in the different catalogues and studies, as listed in Col. 11 and referenced in Table \ref{Table:SEPs_References}. The flare class indicated in square brackets is based on the 1-8 {\AA} channel measurements of the X-Ray telescopes on board GOES. To be consistent with previous statistical studies \citep[e.g.][]{Richardson2014} we used the flare location as the site of the putative source of electrons.

Column 5 in Table \ref{Table:SEP_list} shows the MESSENGER connection angle (CA), which is the longitudinal separation between the flare site location and the footpoint of the magnetic field line connecting to the spacecraft, based on a nominal Parker spiral, as discussed below. Positive CA denotes a flare source located at the western side of the spacecraft's magnetic footpoint. The magnetic footpoint for MESSENGER was estimated assuming a Parker spiral with a constant speed of 400 km s\textsuperscript{-1} using the \textit{Solar-MACH tool}  available online\footnote{\url{https://doi.org/10.5281/zenodo.7100482}\label{Solar-Mach}} \citep{Gieseler2022, Gieseler2022arXiv}, as MESSENGER lacks solar wind measurements. The heliocentric distance of the MESSENGER spacecraft at the time of the event is given in Col. 6, which varies between 0.31 au and 0.47 au during the time interval considered in this study. Column 7 summarizes the 71-112 keV electron peak intensities corresponding to the prompt component of the event as discussed above. The pre-event background level is given in parenthesis. We observed a SEE event on 2012 March 9, but the spacecraft entered in safe mode a few minutes after the onset, so no peak intensity was measured and it was not included in the table. In order to keep the self-consistency of the analysis, events number 6 and 7 measured in August 2011 during the period of increased geometric factor of the MESSENGER/EPS instrument were not included in the study. A detailed description of Cols. 8-10 are given in the following sections. Col. 11 summarizes the references of catalogues and studies that were consulted during the compilation of the list, as detailed in Table \ref{Table:SEPs_References}.  
 
 We found a CME (CME-driven shock) related to the SEE event in 57 (56) events. We indicate with NS next to the event date in Col. 2 of Table \ref{Table:SEP_list} when no CME-driven shock was associated to the SEE event. For these associations we previewed the available conoragraphic data near the flare and SEP onset times and registered the related events \citep[e.g.][]{2009OntiverosVourlidas}. In almost all the cases, the CMEs and CME-driven shock waves were very prominent and clearly related to the flare eruption. In a second study, we will analyse the relations between the electron peak intensities and the properties of the associated parent solar activity (flare, CME, CME-driven shock) for the SEE events measured by MESSENGER.
 
 We consider an event to be widespread when either the MESSENGER |CA| is more than 80$^{\circ}$ or the longitudinal separation between MESSENGER and another spacecraft near 1 au that detected the event was more than 80$^{\circ}$ \citep{Dresing2014}. We indicate these events with (*) next to the event number in Col. 1 of Table \ref{Table:SEP_list}. A total of 44 SEE events were widespread. However, the number of widespread events could be larger since, apart from not sampling all the heliolongitudes with the existing constellation of spacecraft, there were events with a high prior-event-related background, or with no data available for some of the spacecraft, so no increase could be measured.  Relativistic ($\sim$1 MeV) electron intensity enhancements were observed in 37 events, as indicated with an ($\diamond$) in Col. 11 of the list. Thus, the majority of the events detected by MESSENGER are accompanied by a CME and a CME-driven shock, with a high peak intensity level and the presence of $\sim$1 MeV electrons, which are observed by widely separated spacecraft. This type of SEE events is expected due to the high background level of MESSENGER/EPS that prevents the instrument from measuring less intense events \citep[e.g. figure 1 in][]{Lario2013}. 
 
Figure \ref{fig:stats_Int_CA} shows the 71-112 keV electron peak intensities as a function of the CA.
The events with the largest intensities are observed between -80$^{\circ}\lesssim$CA$\lesssim$45$^{\circ}$, 
including the well-connected events at CA$\sim$0$^{\circ}$, with a trend toward negative CA values. Poorly connected events at longitudes CA$\lesssim$-80$^{\circ}$ or CA$\gtrsim$45$^{\circ}$ tend to have intensities below $\sim$10$^{5}$ (cm$^{2}$sr s MeV)$^{-1}$. The highest SEE intensities observed by the MESSENGER mission, showing peaks above 10\textsuperscript{7} (cm\textsuperscript{2} sr s MeV)\textsuperscript{-1}, are SEE events \#5 (2011/06/04) and \#19 (2012/03/07), discussed in detail by \cite{2013Lario_intense}; event \#36 (2013/08/19), studied in detail by  \cite{2021Rodriguez-Garcia}; and event \#53 (2014/09/01).
 \begin{table*}[htb]
\centering
\caption{References for SEP events catalogues and studies.}
\label{Table:SEPs_References}
\small
\begin{tabularx}{1\textwidth}{ccc} 
\hline
\hline
Ref.\#&Paper/list& Reference\\
a & Longitudinal and radial dependence of solar energetic particle intensities: & \cite{Lario2013} \\
&STEREO, ACE, SOHO, GOES, and MESSENGER observations&\\
b &> 25 MeV proton events observed by the High Energy Telescopes on the STEREO A and B &\cite{Richardson2014} \\
& spacecraft and/or at Earth during the first $\sim$ seven years of the STEREO mission\\

c &Solar flares, coronal mass ejections and solar energetic particle events characteristics & \cite{Papaioannou2016}\\
d & Catalogue of >55 MeV wide-longitude solar proton events observed by SOHO& \cite{Paassilta2018} \\
&ACE, and the STEREOs at $\sim$ 1 au during 2009-2016\\
e & Connecting the properties of coronal shock waves with those of solar energetic particles & \cite{Kouloumvakos2019} \\
f &Statistical study on multispacecraft widespread solar energetic particle events during solar cycle 24 &\cite{Xie2019} \\
g & Statistical analysis of the relation between coronal mass ejections and solar energetic particles & \cite{Kihara2020}   \\
 \hline
\end{tabularx}
\begin{flushleft}


\end{flushleft}
\end{table*}

\section{Radial dependence of the peak intensity in SEE events measured by MESSENGER}
\label{sec:radial_dependence}
\begin{figure*}[htbp]
\centering
  \resizebox{\hsize}{!}{\includegraphics{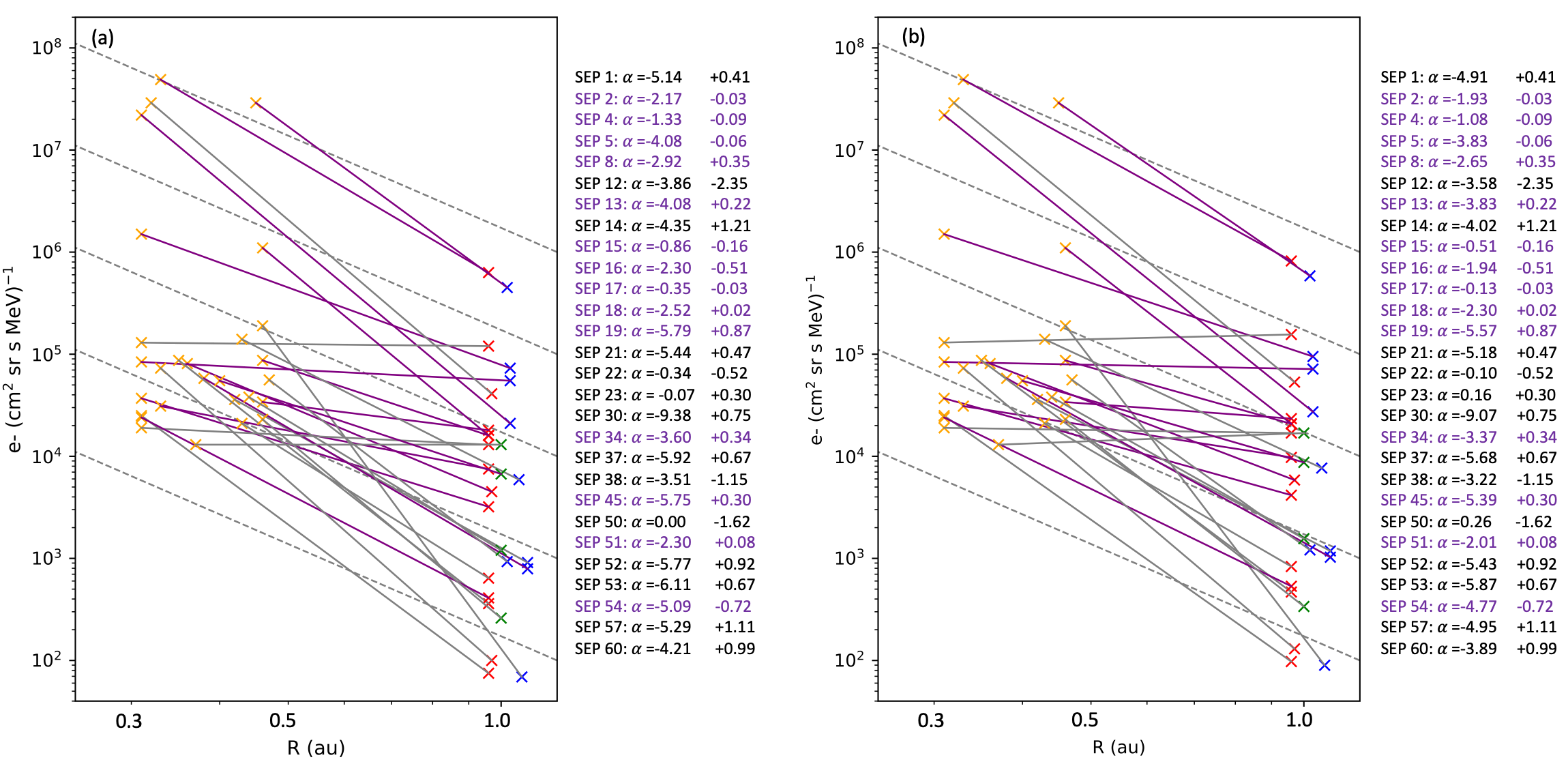}}

     \caption{Radial dependence of the quasi-relativistic electron intensities in SEE events measured by MESSENGER. \textit{(a)} The lines connect the SEE events for which the longitudinal separation of the nominal footpoints of MESSENGER and the respective observation point near 1 au were $\leq$ 35$^{\circ}$. The orange, red, blue and green crosses indicate the peak intensities observed by MESSENGER, STEREO-A, STEREO-B, and ACE, respectively. The dashed lines indicate a R\textsuperscript{-3} radial dependence. The legend on the right indicate the $\alpha$ index
     if a radial dependence $\propto$ R\textsuperscript{$\alpha$} is assumed for each of the SEE events listed. The numbers next to each $\alpha$ index are corrections based on the small longitudinal effect (details given in the text). The purple color indicates the subset of events where the separation of the nominal footpoints of MESSENGER and the respective spacecraft near 1 au were < 20$^{\circ}$. \textit{(b)} Same as in (a) but including an inter-spacecraft calibration factor of 1.3 on the STEREO measurements (details given in the text).   }
     \label{fig:radial_dep_error}
\end{figure*}
\begin{figure*}
\centering
\includegraphics[scale=0.6] {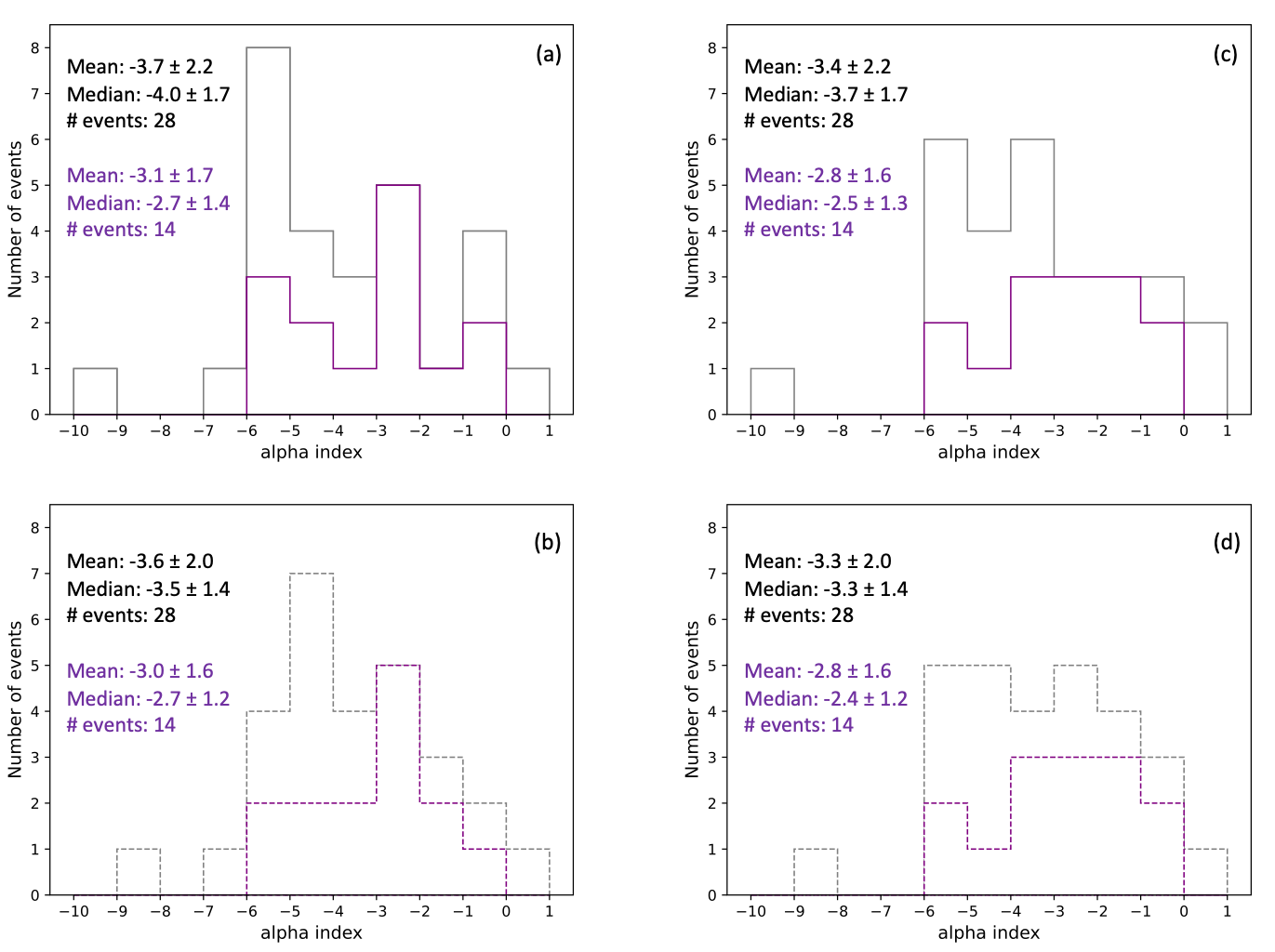}
     \caption{Histograms of the indices $\alpha$ for a radial dependence $\propto$ R\textsuperscript{$\alpha$} of the quasi-relativistic electron intensities in SEE events measured by MESSENGER. \textit{(a)} Gray (purple) color indicates the $\alpha$ indices for those SEP events for which the longitudinal separation of the nominal footpoints of MESSENGER and the respective spacecraft observing near 1 au were $\leq$ 35$^{\circ}$ (< 20$^{\circ}$). \textit{(b)} Same as in (a) but including a correction for the different connection angle. \textit{(c), (d)} The same as in (a) and (b) but including a calibration factor of 1.3 on STEREO measurements. The legend shows both the mean and standard deviation and the median and median absolute deviation. Details given in the text.      }
     \label{fig:alpha_histograms}
\end{figure*}
\begin{table*}[ht]
\centering
\begin{minipage}{\textwidth} 
\small

\caption{Summary of $\alpha$ indices presented in Figs. \ref{fig:alpha_histograms} and \ref{fig:alpha_histograms_solo} for a radial dependence of the peak intensities $\propto$ R\textsuperscript{$\alpha$}}.
\label{table:summary of alpha indices}
\begin{tabularx}{0.95\textwidth}{cccccc} 
\hline
\hline
& &\multicolumn{2}{c}{MESSENGER}&\multicolumn{2}{c}{Solar Orbiter}\\
Ref.&Corrections\hspace{0.1cm}\&\hspace{0.1cm}Subsample&$\alpha$\textsubscript{Med}$\pm$MAD  &<$\alpha$>$\pm$SD &$\alpha$\textsubscript{Med} $\pm$MAD &<$\alpha$>$\pm$SD\\
 \hline
(1)&(2)&(3)&(4)&(5)&(6)\\
 \hline
(a)& No corrections\hspace{0.1cm}\& $\leq$35$^{\circ}$  (<20$^{\circ}$)&-4.0$\pm$1.7 (-2.7$\pm$1.4)&-3.7$\pm$2.2 (-3.1$\pm$1.7)&-2.5$\pm$1.0 (-2.9$\pm$1.3)&-2.4$\pm$2.3 (-2.5$\pm$2.4)\\
(b)& Corrected intensity\hspace{0.1cm}\&  $\leq$35$^{\circ}$(<20$^{\circ}$) &-3.5$\pm$1.4 (-2.7$\pm$1.2)&-3.6$\pm$2.0 (-3.0$\pm$1.6)&-1.5$\pm$1.4 (-1.6$\pm$1.5)&-1.3$\pm$2.6 (-1.4$\pm$2.8)\\
(c)& Inter-cal. factor\hspace{0.1cm}\&  $\leq$35$^{\circ}$(<20$^{\circ}$)&-3.7$\pm$1.7 (-2.5$\pm$1.3)&-3.4$\pm$2.2 (-2.8$\pm$1.6)&-1.6$\pm$0.5 (-1.9$\pm$1.2)&-1.5$\pm$2.3 (-1.6$\pm$2.5)\\
(d) &Corr. int. \& inter-cal.\hspace{0.1cm}\&  $\leq$35$^{\circ}$(<20$^{\circ}$)&-3.3$\pm$1.4 (-2.4$\pm$1.2)&-3.3$\pm$2.0 (-2.8$\pm$1.6)&-0.8$\pm$1.5 (-0.8$\pm$1.5)&-0.5$\pm$2.9 (-0.5$\pm$3.1)\\
 \hline

\hline
\\
\end{tabularx}
\footnotesize{ \textbf{Notes.} Column (1): panel reference in Figs. \ref{fig:alpha_histograms} and \ref{fig:alpha_histograms_solo}. Column (2): corrections included in the measured intensities for the considered subsample, namely for the events where the longitudinal separation of the nominal footpoints of MESSENGER or Solar Orbiter and the respective spacecraft observing near 1 au were $\leq$ 35$^{\circ}$ (< 20$^{\circ}$). Column (3)-(4): respectively, median and median absolute deviation (MAD) and mean and standard deviation (SD) values for the subsample $\leq$35$^{\circ}$(<20$^{\circ}$). Column (5)-(6): same as (3)-(4) but for Solar Orbiter. }

\end{minipage}

\end{table*}
In this section we present the selection of SEE events measured by MESSENGER and analyse the radial dependence of the electron peak intensities. 

\subsection{Data source and SEE event selection criteria}
\label{sec:Radial_depence_data_source}
  Observational studies dealing with the radial dependence of SEE intensities are difficult because both radial and longitudinal effects occur together and cannot be easily disentangled in the data analysis \citep{1983McGuire, Lario2006}. In order to separate the longitudinal and radial effects, we selected the SEE events measured by MESSENGER that were also observed by a spacecraft near 1 au when the nominal magnetic connections of both spacecraft were close in longitude, as detailed below. In order to estimate the magnetic connections we assumed a Parker spiral field configuration with a solar wind speed of 400 km s$^{-1}$ \citep[e.g.][]{2013Lario_intense,2020Joyce}. We limited the study to events with an estimated longitudinal separation of the nominal footpoints between MESSENGER and near 1 au spacecraft of $\leq$ 35$^{\circ}$. This number is the same as the maximum difference chosen by former radial dependence studies \citep[e.g.][]{1983McGuire}. This criterion is fulfilled in 38 events out of 61 SEE events observed by MESSENGER, as marked with (+) or (++) in Col. 1 of Table \ref{Table:SEP_list}. The SEE events marked with (++) are restricted to a separation of <\,20$^{\circ}$, as used in \cite{Lario2006, Lario2013}, that was present in a total of 19 events. The specific magnetic connection difference for each event is indicated in Col. 10 of Table \ref{Table:SEP_list}.
  
  It is possible, however, that the telescopes on board different spacecraft detect a different range of pitch angles, even when they are in close magnetic connection and the telescopes are mounted to scan similar portions of the sky. To minimize this effect when  there is a poor observational coverage of the pitch-angle distribution or the latter is not known, the use of omnidirectional intensities at the different spacecraft is more appropriate for the radial dependence analysis. In the case of MESSENGER, only antisunward observations are available. For the study of the radial dependence, we decided to use telescopes on board the spacecraft near 1 au that point mostly in the sunward direction, or along the nominal Parker spiral direction, and to evaluate the effect of the different viewing directions on the radial dependence of the electron peak intensities (discussed in Sect. \ref{sec:summary_discussion}).
  
  Thus, for near 1 au SEE observations, we used data from the STEREO/SEPT Sun-telescope, pointing sunward 45$^{\circ}$ west from the Sun-spacecraft line; and the ACE/EPAM/Deflected Electron (DE) sensor, part of the Low-Energy Magnetic Spectrometer (LEMS30) telescope, oriented at 30$^{\circ}$ from the spin axis of ACE that points mostly towards the Sun. SEPT measures electrons from $\sim$45 keV to 425 keV and the DE detector of the EPAM instrument measures electrons from $\sim$35 keV to 315 keV. To compare with the 71-112 keV electron channel of MESSENGER, we used the added channels 75-105 keV for SEPT and the channel 53 to 103 keV for ACE.
  
  To find the electron peak intensities near 1 au, we used the same criteria explained in Sect. \ref{sec:Data_sources_SEP_selection_criteria} regarding the prompt component of the peak intensity. As indicated in Col. 9 of Table \ref{Table:SEP_list}, events number 30, 35, 45, 46, 47, 54, 55 and 58 were either affected by ion contamination (in the case of STEREO/SEPT), no data were available, or a maximum peak intensity could not be identified. This was related to either irregular time-intensity profiles, a new SEP injection occurring before a peak intensity could be detected, or the event showing a gradual or continuous increase. We also excluded events number 6 and 7, as explained in Sect. \ref{sec:MESSENGER SEE list}. Therefore, we finally selected 28 events for the radial dependence study, 14 of which present a longitudinal separation of the nominal footpoints of MESSENGER and the respective observing point near 1 au of < 20$^{\circ}$.

\subsection{Radial dependence of the electron peak intensities}
\label{sec:Radial_dependence}

The electron peak intensities measured near 1 au by STEREO or ACE spacecraft are listed in Col. 9 of Table \ref{Table:SEP_list}. The units for both the peak intensity and the pre-event background intensity (in parenthesis) are particles (cm\textsuperscript{2} sr s MeV)\textsuperscript{-1}. The connection angle difference between the respective spacecraft and MESSENGER (CA\textsubscript{near 1au}-CA\textsubscript{MESS}) is  given in Col. 10. Since no SEP data were obtained when MESSENGER was close to Earth before being switched off to start the cruise phase to Mercury, no intercalibration was possible between MESSENGER/EPS intensities and intensities from near-Earth spacecraft. Therefore, in order to compare intensities measured by either STEREO/SEPT or ACE/EPAM/DE with those from  MESSENGER/EPS we adopted the following approach. First, we directly compared MESSENGER/EPS with STEREO/SEPT data without using any scaling correcting factor, but dividing ACE data by the inter-spacecraft calibration factor of 1.3 with STEREO \citep[estimated in Fig. 2 in][]{Lario2013}, as noted in Col. 9 of Table \ref{Table:SEP_list} with (\&). However, in the decay phase of several events, similar electron intensities have been measured between MESSENGER and ACE spacecraft \citep{Lario2011ACEnews}, suggesting the presence of a reservoir effect in which comparable intensities are typically measured between distant spacecraft \citep{1972McKibben, 2010Lario, 1992Roelof}. Based on this, we also included the results when multiplying the STEREO data by an inter-spacecraft calibration factor of 1.3, but without applying any correction to ACE data. As discussed below, the intercalibration factor has little influence on the results, due to the strong dependence of the intensity decrease with the radial distance. 

\begin{figure*}
\centering
  \resizebox{0.9\hsize}{!}{\includegraphics{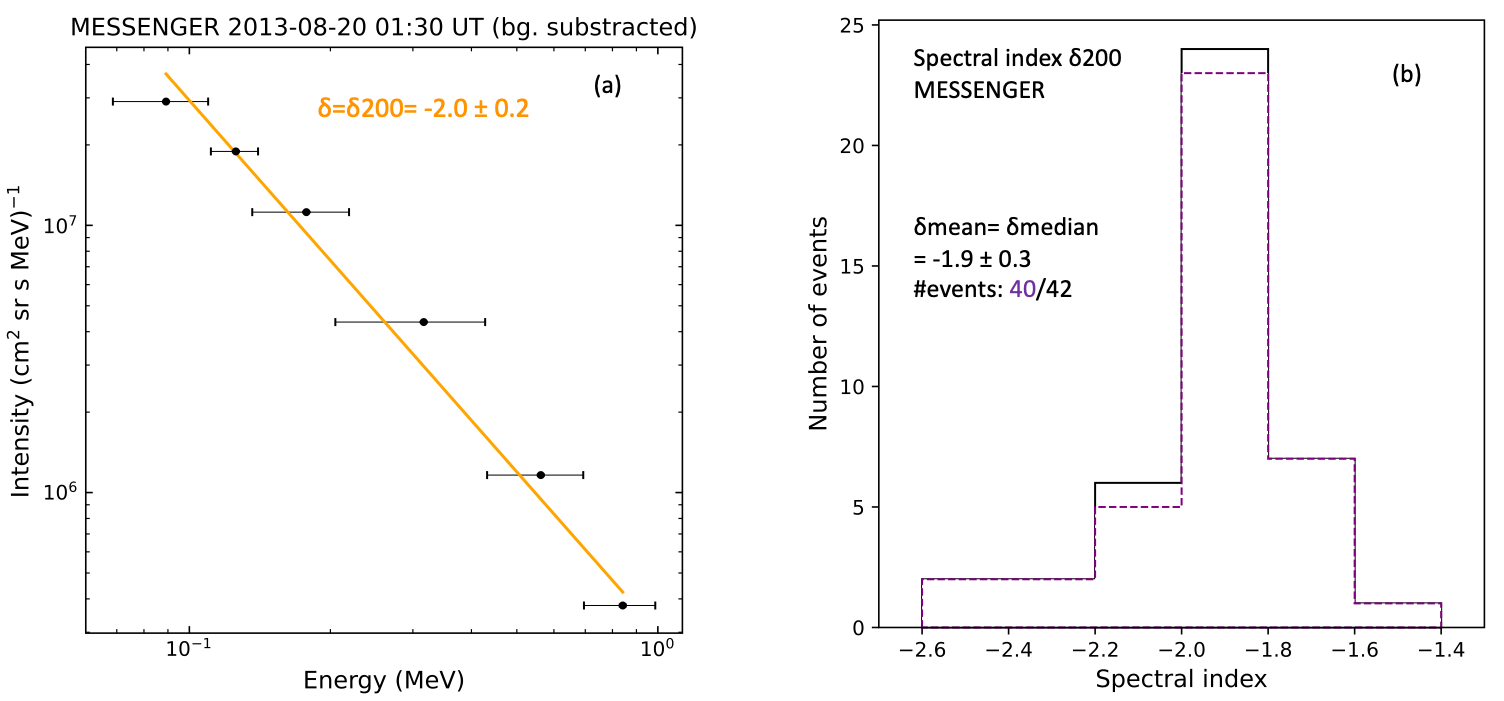}}
     \caption{MESSENGER solar energetic electron event peak spectra. \textit{(a)} Event representative of the peak-intensity energy spectrum, showing a single power-law. The spectral index and its uncertainty are given in the legend. \textit{(b)} Histogram of the spectral indices for the whole sample of events. Mean and standard deviation and median and median absolute deviation values are given in the legend. The purple color indicates the events with a CME-driven shock present. }
     \label{fig:hist_example_spectra}
\end{figure*}
It is possible that the small longitudinal separation between the footpoints of the nominal magnetic field lines connecting to MESSENGER and to near 1 au spacecraft could have an effect on the measured radial dependence of the peak intensities. In order to evaluate this longitudinal effect, we estimated the peak intensity of a hypothetical observer located near 1 au at the same nominal connection angle as MESSENGER, using the longitudinal dependence relation given in equation 3 in \cite{Xie2019}. This formula predicts the 62-105 keV electron intensity observed by a spacecraft near 1 au based on the connection angle to the solar source. For that purpose, we calculated the intensity value for CA=0, I\textsubscript{0}, as:
\begin{equation}\label{eq:gaussian_intensity}
    I\textsubscript{0}=I/exp(-CA\textsuperscript{2}/2\sigma\textsuperscript{2}), where \;
    \sigma = \begin{cases} 7.1 +0.26CA, & \text{if $CA\ge 0$}\\12.9-0.28CA, & \text{if $CA< 0,$} \end{cases}   
\end{equation}
\noindent where I is the peak intensity near 1 au given in Col. 9 of Table \ref{Table:SEP_list}, and CA is the connection angle (in degrees) of the spacecraft observing near 1 au, deduced from Cols. 5 and 10.  Knowing I\textsubscript{0}, from Eq. \ref{eq:gaussian_intensity} we can estimate the `corrected' intensity as I\textsuperscript{'}=I\textsubscript{0}*exp(-CA\textsuperscript{'}\textsuperscript{2}/2$\sigma\textsuperscript{'}$\textsuperscript{2}) using the CA\textsuperscript{'} of MESSENGER. For the sake of brevity, we refer to this intensity corrected by the small difference in the connection angle between near 1 au spacecraft and MESSENGER as the `Corrected intensity' in the rest of the paper. The Corrected intensity is given in squared brackets in Col. 9 of Table \ref{Table:SEP_list}. Then, we used this corrected value to evaluate the pure radial dependence in electron peak intensity between MESSENGER and near 1 au spacecraft for exactly the same nominal CA. We note that the ratio between the measured and Corrected intensities ranged from I/I\textsuperscript{'}=$\sim$0.2 in event \#6 to I/I\textsuperscript{'}$ =\sim$9.4 in event \#12. 

Figure \ref{fig:radial_dep_error} shows the distribution of peak intensities as a function of the heliocentric distance. The gray (purple) lines connect the peak intensities in the prompt component of the events for which the separation of the nominal footpoints of MESSENGER and the respective spacecraft observing near 1 au were $\leq$ 35$^{\circ}$ (< 20$^{\circ}$). The indices $\alpha$, if a radial dependence R\textsuperscript{$\alpha$} is assumed for the maximum intensities, are listed on the right of each plot. The positive and negative numbers next to each $\alpha$ index are the values to add or subtract to the index to correct for the longitudinal effect. We note that the correction is either positive or negative in each event, if the connectivity to the flare location of the hypothetical observer near 1au is improved or worsened with respect to the actual observation, respectively. The black dashed lines indicate a R\textsuperscript{-3} radial dependence, which is an upper limit obtained by the observational and modeling studies described in detail in section 3 of \cite{Lario2013}. Whereas in Fig. \ref{fig:radial_dep_error}a we used directly MESSENGER and STEREO data but applying a dividing intercalibration factor of 1.3 to ACE data, Fig. \ref{fig:radial_dep_error}b shows the same peak intensity radial dependences but using directly MESSENGER and ACE data and applying a multiplying factor of 1.3 on STEREO data, as explained above.  

Figure \ref{fig:alpha_histograms} shows the histogram of the alpha indices for the radial dependence R\textsuperscript{$\alpha$} presented in Fig. \ref{fig:radial_dep_error}. Figure \ref{fig:alpha_histograms}a (b) shows the result based on no inter-spacecraft calibration factor between MESSENGER and STEREO for the peak (Corrected) intensity. Figure \ref{fig:alpha_histograms}c (d) shows the result based on multiplying STEREO data by an intercalibration factor of 1.3 for the peak (Corrected) intensity. ACE data are always inter-calibrated with STEREO data as discussed above.

The legend in Fig. \ref{fig:alpha_histograms} shows the mean and the standard deviation (SD) and the median and the median absolute deviation \citep[MAD;][]{2012Feigelson} of the $\alpha$ index. Based on 28 SEE events, without including the corrections due to the intercalibration factor or to the longitudinal effect the median value of the $\alpha$ index is $\alpha$\textsubscript{Med}=-4.0$\pm$1.7. The corrections in the median $\alpha$ index due to the intercalibration factor and longitudinal effect are $\sim$8\% ($\alpha$\textsubscript{Med}=-3.7$\pm$1.7) and $\sim$12\% ($\alpha$\textsubscript{Med}=-3.5$\pm$1.4), respectively. When correcting both effects, the median value is $\alpha$\textsubscript{Med}=-3.3$\pm$1.4. In the case of the reduced sample with |CA difference| < 20$^{\circ}$ (14 SEE events), the median of the $\alpha$ index is $\alpha$\textsubscript{Med}=-2.7$\pm$1.4. In this subsample, there is no influence due to the longitudinal effect on the median value ($\alpha$\textsubscript{Med}=-2.7$\pm$1.2), although the median absolute deviation is lower when correcting for this effect. The correction due to the intercalibration factor is similar to that of the whole sample ($\sim$8\%), with a median $\alpha$ index $\alpha$\textsubscript{Med}=-2.5$\pm$1.3. When considering both corrections, the median $\alpha$ index is $\alpha$\textsubscript{Med}=-2.4$\pm$1.2. A summary of the $\alpha$ indices for the different corrections and subsamples is presented in Cols. (3)-(4) of Table \ref{table:summary of alpha indices}.  
\section{Peak-intensity energy spectra and their radial dependence in SEE events measured by MESSENGER }
\label{sec:peak_spectra}
In this section we present the selection of SEE events measured by MESSENGER where the energy spectra could be determined. We also show the spectra obtained from near 1 au spacecraft when in close
nominal magnetic connection with MESSENGER and study the radial dependence of the electron energy spectrum measured at the peak of the event. 

\subsection{MESSENGER peak spectra}
\label{sec:MESSENGER_peak_spectra}

\begin{figure*}
\centering
  \resizebox{1.0\hsize}{!}{\includegraphics{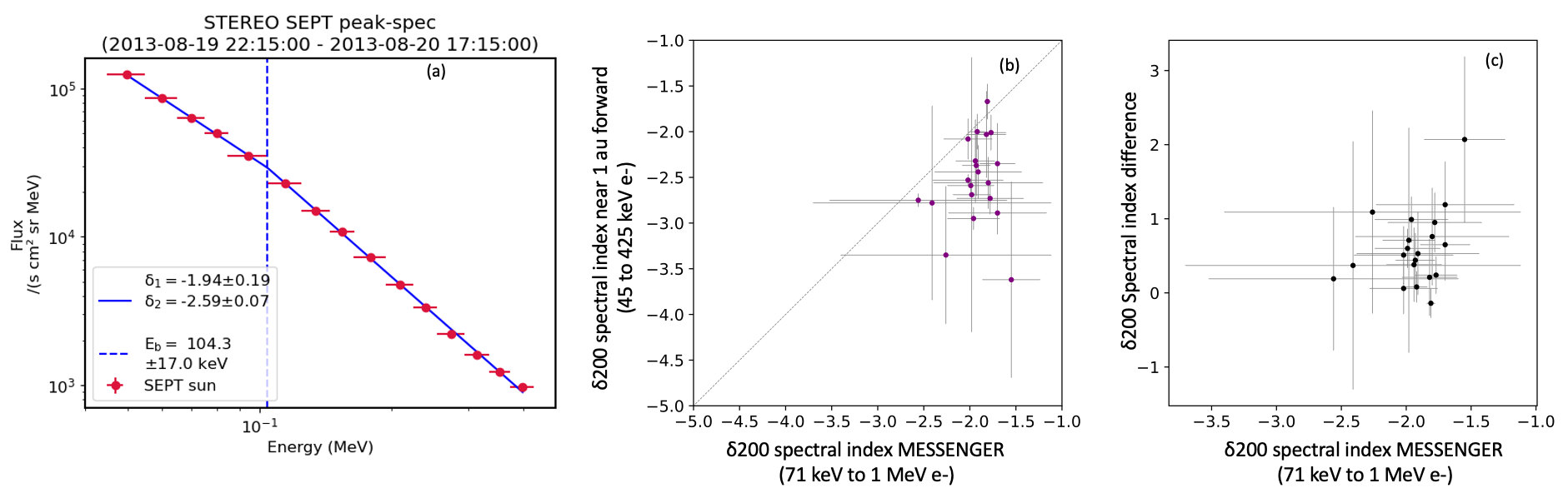}}
     \caption{Comparison between MESSENGER and near 1 au forward spectra. \textit{(a)} Event representative of the STEREO peak-intensity energy spectrum showing a 
     broken power-law. The legend shows the values of the fit parameters: the spectral index below ($\delta$\textsubscript{1}) and above ($\delta$\textsubscript{2}) the spectral transition; E\textsubscript{b} (vertical dashed line). \textit{(b)} Spectral indices near 1 au against the spectral indices at MESSENGER. The dashed line indicates equal indices. \textit{(c)} SEE spectral indices difference ($\delta$200\textsubscript{MESS}-$\delta$200\textsubscript{near\_1au\_forw}) against the spectral index calculated with MESSENGER measurements ($\delta$200\textsubscript{MESS}). }
     \label{fig:diff_spectra_sun}
\end{figure*}

We analysed the spectra of the SEE events measured by the MESSENGER mission listed in Table \ref{Table:SEP_list}. The EPS instrument measured electrons from $\sim$20 to $\sim$1000 keV in 10 energy channels, mainly in the anti-Sun direction. The first four bins could not be used due to instrumental effects, so the energies used in this analysis are from $\sim$71 keV to $\sim$1 MeV divided into six energy bins. For each one of the events, we took the time-of-maximum (TOM) based on the 71-112 keV channel using one-hour averages to increase the statistics (Col. 7 in Table \ref{Table:SEP_list}), and read the intensity at this time for the rest of the energy channels. We subtracted the pre-event background for each energy channel, which includes background increase caused by preceding events. We did not separate or discard events depending on their rise times, intensities, delays or correlations to solar flares, CMEs or type III radio bursts. 

The criteria used for the spectral fitting are as follows. (1) The spectrum should include at least four energy bins. (2) An energy bin is eliminated if it shows similar or higher intensities than its lower energy neighbour (e.g. due to ion contamination), which would correspond to a power-law with a positive slope. (3) The relative uncertainty of the electron peak intensity with the background subtracted should be below 50\%. Following these criteria, we had to discard 19 events from the original list of 61 SEE events. 

For all the remaining 42 events, we fitted the spectrum allowing for either a broken or a single power-law shape \citep[e.g.][]{Dresing2020, Strauss2020}, but we found that a single power-law was appropriate to fit the whole sample of 42 events. Thus, no energy transition was found for the selected events. Figure \ref{fig:hist_example_spectra}a shows a representative peak-intensity energy spectrum, with a fit resembling a single power-law.  The resulting spectral index and its uncertainty is given in the legend, where the uncertainty is calculated for a confidence interval of 95\%. We note that the reduced number and large width of the energy channels might be behind the single power-law fitting, but the visual impression of the spectrum is consistent with a transition near 300 keV. 

Figure \ref{fig:hist_example_spectra}b shows the histogram of the spectral indices for the 42 SEE events, where the mean and median values coincide <$\delta$>=$\delta$\textsubscript{Med}= -1.9 $\pm$ 0.3. The purple subsample corresponds to the events with the presence of a CME-driven shock (40 out of 42 events), as observed by the EUV and white-light images from STEREO and SOHO points of view. This subsample presents the same mean and median values for the spectral index as the whole sample. 

As discussed above for the representative example, the absence of a broken power-law could be due to several factors, such as the large width and small number of the energy bins provided by MESSENGER/EPS, and the adoption of energy bins only above $\sim$70 keV for the fitting. Due to this instrumental limitation in finding potential spectral transitions, we also fitted the sample using only the three highest energy bins that observed the increase in intensity over the background level ($\sim$200 keV to $\sim$1 MeV). We note that these spectral indices have higher uncertainties due to the low number of energy bins used for the fitting. However, we observed that these fittings were consistently softer than those using all bins available. This could be related to the potential existence of spectral transitions, which can not be fitted by the model due to insufficient energy resolution. We note that the limitations discussed above to calculate the energy spectra, regarding the anti-Sun pointing of MESSENGER, or the number and width of the energy bins, are not present in new ongoing missions, such as Solar Orbiter, which allows the exploration of spectral transitions near 0.3 au, as discussed in Sect. \ref{sec:SolO peak spectra}.  

\begin{figure*}
\centering
  \resizebox{1.0\hsize}{!}{\includegraphics{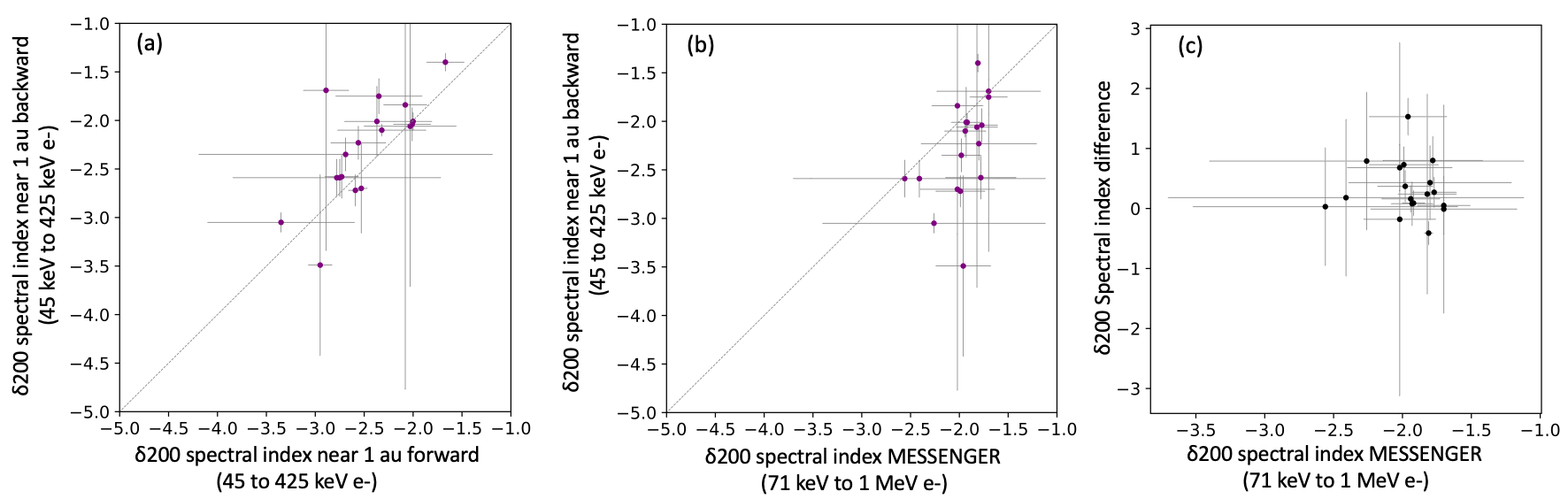}}
     \caption{Comparison between backward and forward spectra. \textit{(a)} Spectral indices near 1 au: backward spectra against forward spectra.  \textit{(b)} and \textit{(c)} The same as in Figs. \ref{fig:diff_spectra_sun}b and \ref{fig:diff_spectra_sun}c but for the backward spectra. }
     \label{fig:diff_spectra_asun}
\end{figure*}
 
\subsection{Spectra comparison: from 0.3 au to 1 au}
\label{sec:spectra_comparison}

In this section we present the comparison of the energy spectra measured by MESSENGER and by spacecraft near 1 au. 
We limited the study to events with a longitudinal separation of the nominal footpoints of MESSENGER and the respective observations near 1 au of $\leq$ 35$^{\circ}$ (marked with + and ++ in Table \ref{Table:SEP_list}, as discussed in Sect. \ref{sec:Radial_depence_data_source}). This subsample includes 18 (2) events measured by STEREO/SEPT (Wind/3DP). Because of the anti-Sun pointing of the MESSENGER/EPS instrument, we adopted the following approach. We compare the spectral indices of MESSENGER with the forward spectra measured near 1 au in Sect. \ref{subsubsection:backward_versus forward} and with the backward spectra measured near 1 au in Sect. \ref{subsubsection:backward_versus_backward_spectra} to evaluate the radial dependence of the spectral indices in both configurations. We refer as forward (backward) spectra as the energy spectra calculated using the flux of particles mainly coming from the Sun (anti-Sun) direction.

Thus, we used the forward and backward spectra obtained from near 1 au spacecraft, where the process followed to determine the spectra of STEREO is similar as that used by \cite{Strauss2020}. In this case we used hourly averages to compare with MESSENGER/EPS measurements. The electron STEREO/SEPT bins contaminated by ions were removed from the study if the estimated ion contamination was higher than 40\%. In the case of Wind/3DP, the instrument measures energetic electrons from 26 keV to 522 keV binned in eight different pitch angles. We selected the pitch angle covering the flux of particles mainly coming from the Sun or anti-Sun direction according to the magnetic field polarity and removed the first noisy channel to obtain the spectrum. To be consistent, we also used hourly averages. Ion contamination did not affect the peak of the SEE events measured by Wind presented in this sample. In the case of the particles coming mainly from the anti-Sun direction, there were two events where the increase was not measured above the background level of Wind/3DP instrument that were therefore excluded of the study. Thus, we only included the 18 events measured by STEREO in the backward spectra analysis. To be able to compare the single power-law indices from MESSENGER with the broken power-law indices from near 1 au spacecraft, we chose the $\delta$200 index for the three spacecraft, namely MESSENGER, STEREO and Wind. The term $\delta$200 was introduced by \cite{Dresing2020}, where $\delta$200 corresponds to the spectral indices found in the energy range around 200 keV  \citep[equation 4 by][]{Strauss2020}. Then, in the case of single-power-law events $\delta$ and $\delta$200 have equal values.

\subsubsection{Backward spectra near 0.3 au versus forward spectra near 1 au}\label{subsubsection:backward_versus forward}
Figure \ref{fig:diff_spectra_sun}b shows the forward spectral index near 1 au against the spectral index at MESSENGER. We note that MESSENGER spectral index variation is smaller than that of the spectra obtained from 1 au measurements, whereas the error bars are similar at both locations. Almost all of the points (19 out of 20) lie below the unity line. We note that the spectral indices are negative numbers and being below the unity line means that the spectral indices at MESSENGER are larger (closer to -1) than those near 1 au. The median and MAD values for the ratio $\delta$200\textsubscript{MESS}/$\delta$200\textsubscript{near\_1au\_forw} are 0.8$\pm$0.1. This means that, on average, the MESSENGER backward spectra are $\sim$20\% harder than the forward spectra near 1 au. 

Figure \ref{fig:diff_spectra_sun}c presents the difference of these indices, showing that near 1 au and above 200 keV the spectra are softer than near 0.3 au, namely the spectral index difference between MESSENGER and STEREO or Wind ($\delta$200\textsubscript{MESS}-$\delta$200\textsubscript{near\_1au\_forw}) is always positive within the error bars. The softening in the spectra might be related to IP scattering processes. As an example, Fig. \ref{fig:diff_spectra_sun}a shows the peak spectra obtained from STEREO/SEPT measurements for the event on 2013 August 19 to be compared with the peak spectra obtained from MESSENGER/EPS measurements shown in Fig. \ref{fig:hist_example_spectra}a. This event was studied in detail by \cite{2021Rodriguez-Garcia} who interpret the observations in terms of strong scattering present between the locations of MESSENGER and STEREO-A which may be the cause of the spectral softening. The spectral index of $\delta$200=-2.0$\pm$0.2 at MESSENGER, softens to $\delta$200=-2.59$\pm$0.07 near 1 au.

\subsubsection{Backward spectra near 0.3 au versus backward spectra near 1 au}\label{subsubsection:backward_versus_backward_spectra}

Figure \ref{fig:diff_spectra_asun}b shows the backward spectral index near 1 au against the spectral index at MESSENGER. Similarly to Fig. \ref{fig:diff_spectra_sun}b, The majority of the points (15 out of 18) lie below the unity line.  The median and MAD values for the ratio $\delta$200\textsubscript{MESS}/$\delta$200\textsubscript{near\_1au\_back} are 0.9$\pm$0.1. This means that, on average, the MESSENGER backward spectra is $\sim$10\% harder than the backward spectra near 1 au.

Figure \ref{fig:diff_spectra_asun}c presents the difference of these indices as a function of the spectral index at MESSENGER, showing that near 1 au and above 200 keV the backward spectra is also softer than near 0.3 au, namely the spectral index difference between MESSENGER and STEREO ($\delta$200\textsubscript{MESS}-$\delta$200\textsubscript{near\_1au\_back}) is always positive within the error bars. Fig. \ref{fig:diff_spectra_asun}a shows that in most of the events (13 out of 18) the spectra near 1 au are harder for the backward-scattered population than for the anti-sunward propagating beam, which is in agreement with the results by \cite{Strauss2020}. The ratio of the spectral index $\delta$200\textsubscript{forward}/$\delta$200\textsubscript{backward} using data from STEREO is 1.1$\pm$0.1. It means that near 1 au the forward spectra is $\sim$10\% softer than the backward spectra.


\begin{figure*}
\centering
  \resizebox{0.93\hsize}{!}{\includegraphics{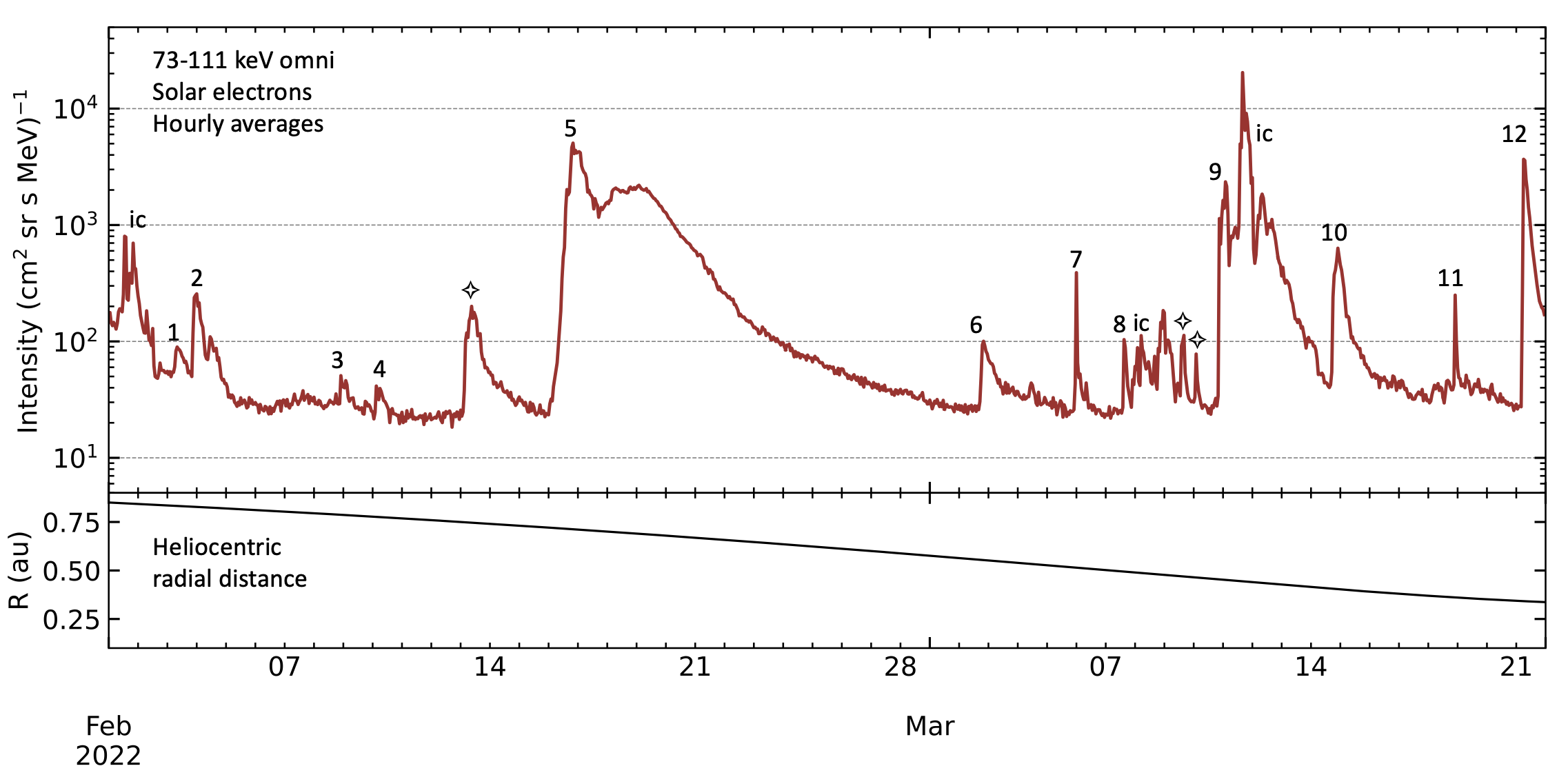}}
     \caption{Hourly averages of 73-111 keV electron omnidirectional intensities measured by EPD/EPT on board Solar Orbiter (top panel). The time interval covers from 2022 February 1 to 2022 March 22 when Solar Orbiter heliocentric distance varied between 0.34 and 0.83 au (bottom panel). The numbers 1 to 12 indicate the SEE events included in the study (details given in  the text).  }
     \label{fig:solo_data}
\end{figure*}
  \begin{figure}
  \resizebox{1.0\hsize}{!}{\includegraphics{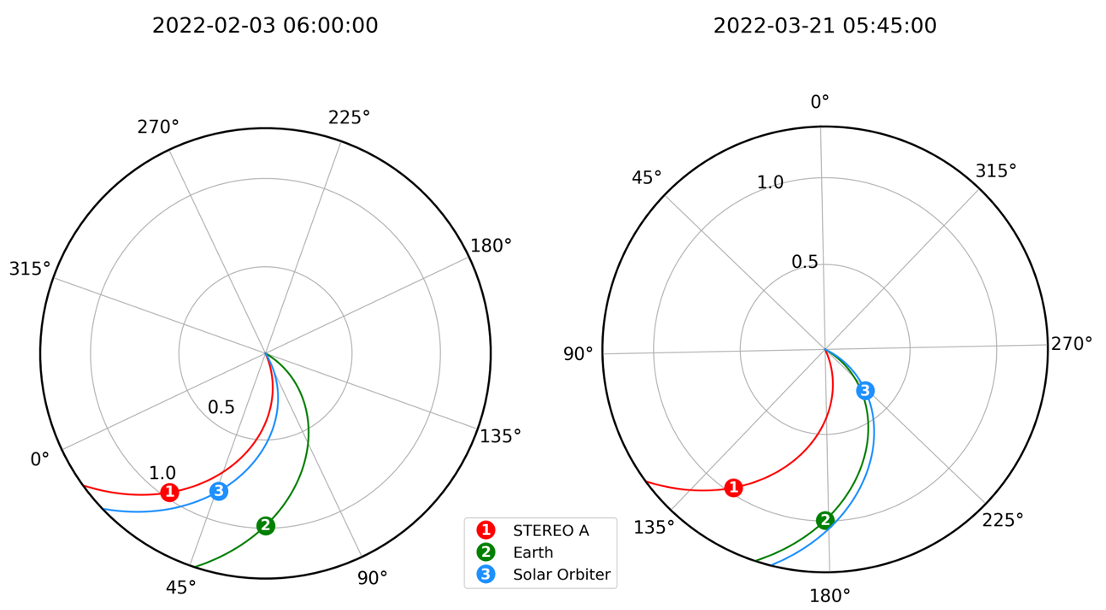}}
     \caption{Configuration of spacecraft in the heliosphere on 2022 February 3 (left) and March 21 (right), close to the Solar Orbiter first perihelion passage. Numbered symbols indicate the observers’ locations and the spiral lines the corresponding magnetic field lines connecting them to the Sun using a speed of 400 km s\textsuperscript{-1}. Radial distance and angular information given in au and Carrington longitude, respectively.  Source: Solar-MACH (\url{https://solar-mach.github.io/}).  }
     \label{fig:solar_mach}
\end{figure} 


\section{SEE events measured by Solar Orbiter near its first close perihelion passage}
\label{sec:solar_orbiter_see_events}
In this section we analyse 12 SEE events measured by Solar Orbiter from 2022 February 1 to March 22, when the spacecraft was close to its first perihelion passage, in particular reaching a distance of 0.32 au on 2022 March 26. During the time of analysis, coinciding with the rising phase of solar cycle 25, Solar Orbiter's radial distance varied from 0.34 to 0.83 au.

 Left (right) panel of Fig. \ref{fig:solar_mach} shows the locations in the heliosphere of Solar Orbiter, STEREO, and near Earth spacecraft at the beginning (end) of the period of analysis. Thus, the site of the nominal magnetic connection of Solar Orbiter on the Sun was very close to those of near 1 au spacecraft such as STEREO-A, ACE and Wind. This configuration of spacecraft gives us the opportunity of studying the radial dependencies of the electron peak intensities and spectral indices, and comparing them with the results obtained from the analysis using MESSENGER data.

\subsection{Solar Orbiter data source and SEE event selection criteria}
\label{sec:Solo_data_source}

The EPT instrument on board Solar Orbiter measures electrons with energies from $\sim$30 keV to $\sim$470 keV divided into 34 energy bins. The electron bins used here for the identification of the SEE events cover the energy range between 73 and 111 keV. These energies are similar to the 71-112 keV energy channel used in the case of MESSENGER. The EPT instrument consists of two double-ended telescopes, where EPT-1 is pointing sunward and anti-sunward along the nominal Parker spiral, and EPT-2 is pointing northward and southward with some inclination \cite[as shown in the figure 4 by][]{Rodriguez-Pacheco2020}. In this study we used the average of the intensity measured by the four telescopes, namely omnidirectional intensities. The reason for using the omnidirectional averaging is to compare with the STEREO-A measurements, where the nominal pointing directions of the STEP telescopes changed in 2015, as discussed in Sect. \ref{sec:SolO_radial_dependence}.

The top panel of Fig. \ref{fig:solo_data} shows the 73-111 keV electron omnidirectional intensities measured by EPD/EPT on board Solar Orbiter from 2022 February 1 to March 22, where the SEE events appear as vertical increases. We used hourly averages of the particle intensities to be consistent with the analysis done using MESSENGER data. The periods affected by ion contamination are indicated with `ic' in the figure. The bottom panel shows the variation of the heliocentric distance of Solar Orbiter during the time of study. The criteria used to select the SEE events are similar to those in Sect. \ref{sec:Data_sources_SEP_selection_criteria}. In this case, we only selected the events measured by Solar Orbiter which were also clearly identified by eye above the background level in STEREO-A or ACE measurements. We note that Solar Orbiter is able to measure electron intensities well below $\sim$10\textsuperscript{4} (cm\textsuperscript{2} sr s MeV)\textsuperscript{-1} due to the lower background level of the EPD/EPT instrument in comparison with the MESSENGER/EPS instrument in the energy range studied here, as shown in Fig. \ref{fig:MESS_mission_71-112keV}a. 

\subsection{Solar Orbiter SEE event list}
\label{sec:Solo_event_list}

Table \ref{Table:SEP_list_Solo} shows the list of the 12 SEE events selected here. These events are indicated with numbers 1 to 12 in Fig. \ref{fig:solo_data}. There were several SEE events measured by Solar Orbiter during the period of analysis that were not included in the list following the criteria discussed in Sects. \ref{sec:Data_sources_SEP_selection_criteria} and \ref{sec:Solo_data_source}. Three of these events are indicated with a diamond in Fig. \ref{fig:solo_data}. There were several injections around the time of the event at the beginning of the day 2022/02/13 and we could not identify a unique and clear solar source. The two small increases measured by Solar Orbiter in the middle of the day 2022/03/09 and the beginning of the day 2022/03/10 were not included as the electron increase measured at STEREO-A were within its background level for both events.

\onecolumn
\small
\begin{landscape}
\begin{ThreePartTable}
\begin{TableNotes}

\item \footnotesize{Columns 1 and 2: Event number and date. Column 3: Type III radio burst onset time. Column 4: Flare location in Stonyhurst coordinates and flare class based on GOES Soft X-ray (SXR) peak flux. Column 5: Longitudinal separation between the flare location and the footpoint of the magnetic field line connecting to Solar Orbiter, based on a 400 km s$^{\ -1}$ Parker spiral (positive connection angle (CA) denotes a flare source located at the western side of the spacecraft magnetic footpoint). Column 6: Solar Orbiter radial distance from the Sun. Column 7: 73-111 keV electron peak intensity measured by Solar Orbiter (SolO). The pre-event background level is shown in parenthesis. Columns 8 and 9: Respectively, Solar Orbiter spectral indices found in the energy range around 70 keV and 200 keV, based on peak electron intensities. Column 10: 75 to 105 keV electron peak intensity measured by near 1 au spacecraft (STA: STEREO-A; STB: STEREO-B), followed in square brackets by the Corrected peak intensity of a hypothetical 1 au observer with exactly the same CA as Solar Orbiter. The pre-event background level is given in parenthesis. Column 11: Name of the 1 au spacecraft and the CA difference (CA\textsubscript{near 1au}-CA\textsubscript{SolO}).  * in Col. 1: Widespread SEP event: Solar Orbiter |CA| or the absolute value of the CA difference with near 1 au spacecraft is $\geq$80$^{\circ}$. \textsuperscript{++} in Col. 1: The absolute value of the difference between the CA of Solar Orbiter and near 1 au spacecraft is < 20$^{\circ}$. \textsuperscript{+} in Col. 1: The absolute value of the difference between the CA of Solar Orbiter and near 1 au spacecraft is: 20$^{\circ}$ $\leq$  |CA difference| $\leq$ 35$^{\circ}$. \textsuperscript{NS} in Col. 2: No CME-driven shock associated to the SEE event.
\textsuperscript{a} in Col. 4: Flare location based on longitude and latitude of the 3D CME apex (not shown).
\textsuperscript{b} in Col. 4: Gradual increase of the X-Ray flux, with a peak at 21:00 UT.
\textsuperscript{\&} in Col. 10: ACE EPAM DE 53-103 keV electron intensity divided by an inter-calibration factor of 1.3.}

\end{TableNotes}

\begin{longtable}{ccccccccccc}

\caption{Solar energetic electron events measured by Solar Orbiter near its first close perihelion.  }\
\label{Table:SEP_list_Solo}\\
\hline
\hline
\multicolumn{4}{c}{Solar event}& \multicolumn{7}{c}{SEE event (based on omnidirectional data)}\\
\cline{2-3}  \cline{5-11}
\#&Date &T-III &Flare&CA &R& I\textsubscript{max\_SolO} (bg)&$\delta$70&$\delta$200&I\textsubscript{max\_near\_1au}[I\textsuperscript{'}\textsubscript{corr.}](bg)& s/c: CA diff. \\
& & onset& loc [class]& && 73 to 111 keV e&SolO&SolO&75 to 105 keV e \\
& &(UT $\pm$ 5 min)&(deg)&(deg)&(au)&(cm\textsuperscript{2} sr s MeV)\textsuperscript{-1}&(-)&(-)&(cm\textsuperscript{2} sr s MeV)\textsuperscript{-1}\\
 \hline
(1)&(2)&(3)&(4)&(5)&(6)&(7)&(8)&(9)&(10)&(11)\\
 \hline
 \endfirsthead
 
\caption{(Continued.)}\\
\hline
\hline
\multicolumn{4}{c}{Solar event}& \multicolumn{7}{c}{SEE event (based on omnidirectional data)}\\
\cline{2-3}  \cline{5-11}
\#&Date &T-III &Flare&CA &R& I\textsubscript{max\_SolO} (bg)&$\delta$70&$\delta$200&I\textsubscript{max\_near\_1au}[I\textsuperscript{'}\textsubscript{corr.}](bg)& s/c: CA diff. \\
& & onset& loc [class]& && 73 to 111 keV e&SolO&SolO&75 to 105 keV e \\
& &(UT $\pm$ 5 min)&(deg)&(deg)&(au)&(cm\textsuperscript{2} sr s MeV)\textsuperscript{-1}&(-)&(-)&(cm\textsuperscript{2} sr s MeV)\textsuperscript{-1}\\
 \hline
(1)&(2)&(3)&(4)&(5)&(6)&(7)&(8)&(9)&(10)&(11)\\
 \hline
 \endhead
\multicolumn{11}{c}{(Continued on next page.)}

\endfoot

\insertTableNotes  
\endlastfoot

1\textsuperscript{++} &\textsuperscript{NS}2022/02/03 &05:55&N16W052 [B9.9]&21 &0.83&8.9$\times$10\textsuperscript{1} (5.4$\times$10\textsuperscript{1})&-&-&5.3$\times$10\textsuperscript{1} [8.7$\times$10\textsuperscript{1}] (3.0$\times$10\textsuperscript{1})& STA: +7$^{\circ}$ \\
2\textsuperscript{++} &\textsuperscript{NS}2022/02/03 &20:40&N16W059 [C1.4]& 28&0.83 &2.4$\times$10\textsuperscript{2} (5.0$\times$10\textsuperscript{1}) &-3.06$\pm$0.19&-& 3.2$\times$10\textsuperscript{2} [4.9$\times$10\textsuperscript{2}] (2.1$\times$10\textsuperscript{1})&STA: +7$^{\circ}$\\
3\textsuperscript{++}&\textsuperscript{NS}2022/02/08 &21:35&S20W045 [C5.2]&15 &0.79  & 5.1$\times$10\textsuperscript{1} (2.9$\times$10\textsuperscript{1})  & -3.60$\pm$0.50&-& 2.5$\times$10\textsuperscript{1} [3.7$\times$10\textsuperscript{1}] (1.2$\times$10\textsuperscript{1}) &STA: +5$^{\circ}$\\
4\textsuperscript{++} &\textsuperscript{NS}2022/02/10 &02:45&S17W056 [B9.3]&27 &0.78  &4.1$\times$10\textsuperscript{1} (2.2$\times$10\textsuperscript{1}) &-& &1.0$\times$10\textsuperscript{1} [1.3$\times$10\textsuperscript{1}] (6.2$\times$10\textsuperscript{0})&STA: +4$^{\circ}$ \\
*5\textsuperscript{++}&2022/02/15 &21:55&N33E134\textsuperscript{a} [-]& -161 &0.72&2.0$\times$10\textsuperscript{3} (2.4$\times$10\textsuperscript{1})&-1.84$\pm$0.25&-2.09$\pm$0.07& Several events mixed &STA: +2$^{\circ}$ \\
6\textsuperscript{++} &\textsuperscript{NS}2022/03/02 &17:45&N16E030 [M2.0]& -57 &0.56&1.0$\times$10\textsuperscript{2} (2.6$\times$10\textsuperscript{1})  &- &-& 1.0$\times$10\textsuperscript{2} [9.4$\times$10\textsuperscript{1}] (1.2$\times$10\textsuperscript{1})&STA: +2$^{\circ}$  \\
7\textsuperscript{++}&\textsuperscript{NS}2022/03/05 &23:55&S15W028 [C1.4]&-1 &0.51&3.9$\times$10\textsuperscript{2} (4.3$\times$10\textsuperscript{1})  &-5.26 $\pm$0.21&-& 2.5$\times$10\textsuperscript{1} [2.7$\times$10\textsuperscript{1}] (9.5$\times$10\textsuperscript{0})& STA: +4$^{\circ}$  \\
*8\textsuperscript{++}&\textsuperscript{NS}2022/03/07 &15:00&S50E090 [C1.1]& -121 &0.49&1.0$\times$10\textsuperscript{2} (2.7$\times$10\textsuperscript{1})  &-3.45$\pm$0.21 &-&1.8$\times$10\textsuperscript{1}[1.6$\times$10\textsuperscript{1}](9.4$\times$10\textsuperscript{0})&STA: +6$^{\circ}$  \\
9\textsuperscript{++}&2022/03/10 &19:10&N25W020 [C2.5\textsuperscript{b}]& -15 &0.45&1.1$\times$10\textsuperscript{3} (3.4$\times$10\textsuperscript{1})  &-2.58$\pm$0.58&-3.89$\pm$0.77&Several events mixed &STA: +10$^{\circ}$\\
10\textsuperscript{++}&2022/03/14 &17:20&S25W090 [B8.5]& 47 &0.41&4.1$\times$10\textsuperscript{2} (4.4$\times$10\textsuperscript{1})  &-1.73$\pm$0.12&-1.73$\pm$0.12&Several events mixed &STA : +17$^{\circ}$\\
11\textsuperscript{+}&\textsuperscript{NS}2022/03/18 &22:15&N20W041 [B2.9]& -15 &0.36&2.5$\times$10\textsuperscript{2} (4.3$\times$10\textsuperscript{1})  &-2.53$\pm$0.60&-&5.1$\times$10\textsuperscript{1} [8.8$\times$10\textsuperscript{1}] (1.8$\times$10\textsuperscript{1}) &STA: +30$^{\circ}$\\
*12\textsuperscript{++}&2022/03/21 &05:40&S35W146\textsuperscript{a} [-]& 81 &0.34&3.7$\times$10\textsuperscript{3} (2.8$\times$10\textsuperscript{1})  &-1.77$\pm$0.49&-2.96$\pm$1.22&5.7$\times$10\textsuperscript{2} [6.3$\times$10\textsuperscript{2}] (8.5$\times$10\textsuperscript{0}) &ACE\textsuperscript{\&} : +4$^{\circ}$\\

 \hline

\end{longtable}

\end{ThreePartTable}
\end{landscape}

\twocolumn 
\begin{figure*}
\centering

  \resizebox{1.0\hsize}{!}{\includegraphics[scale=0.7]{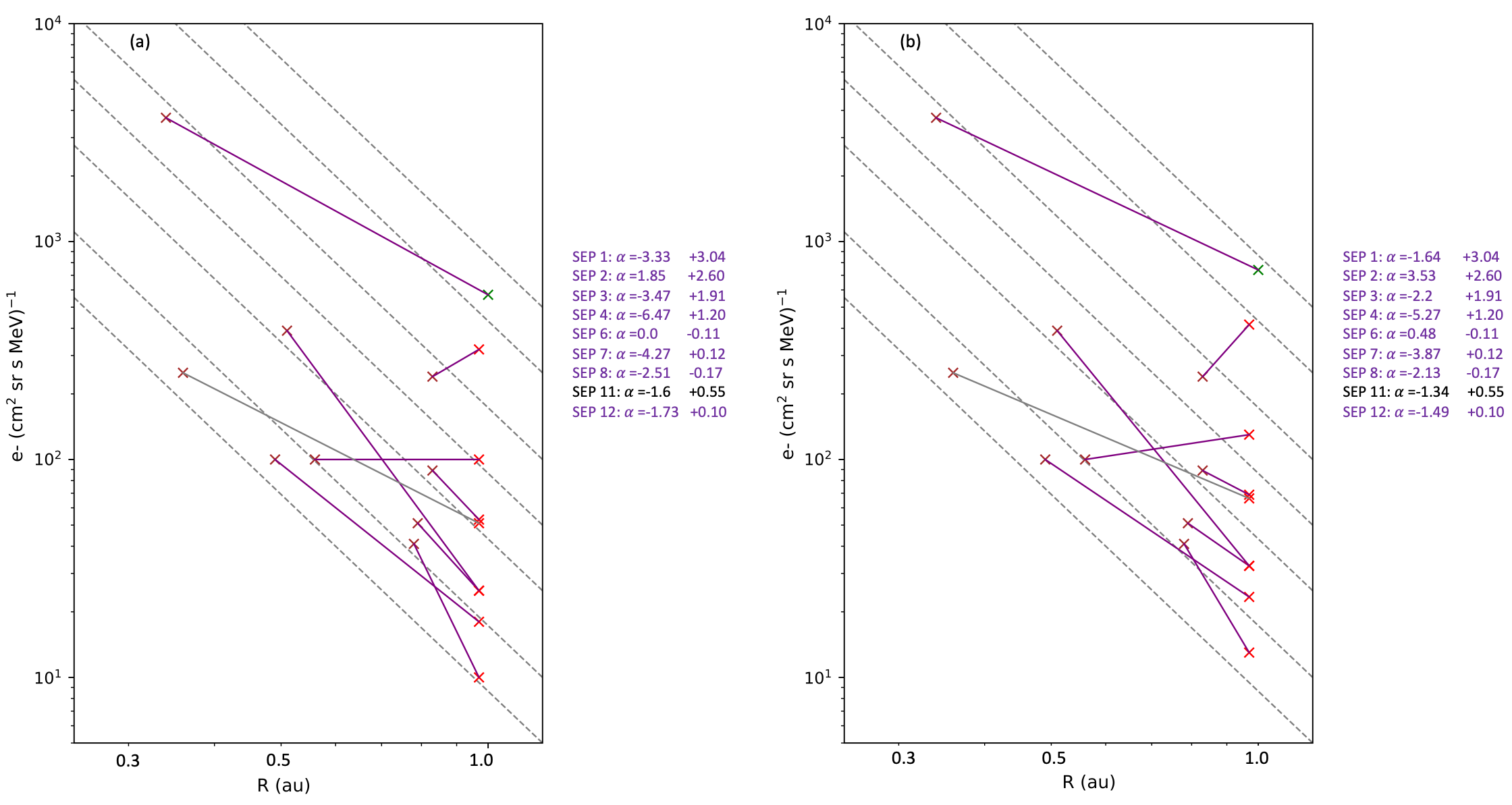}}
     \caption{Radial dependence of the near-relativistic electron intensities in SEE events measured by Solar Orbiter. (\textit{b}) same as in (\textit{a}) but including an inter-spacecraft calibration factor of 1.3 on the STEREO measurements. Details given in Fig. \ref{fig:radial_dep_error} and in the text. }
     \label{fig:solo_radial_depend}
\end{figure*}

The description of Cols. 1-7 in Table \ref{Table:SEP_list_Solo} are the same as in Table \ref{Table:SEP_list}, but related to Solar Orbiter measurements. Columns 8-11 are described in the following sections.  We note that the sample of events included in the Solar Orbiter list is smaller than the list of MESSENGER SEE events, mainly due to the shorter period of analysis in the Solar Orbiter sample. The specific (ascending) phase of the solar cycle analysed here, and the much lower background level of the EPD/EPT instrument on board Solar Orbiter are also some factors related to the difference between the two samples. 

In the Solar Orbiter list we mainly find small events (9 out of 12), with peak 73-111 keV electron intensities that vary from $\sim$10\textsuperscript{1} to  $\sim$10\textsuperscript{3} (cm\textsuperscript{2} sr s MeV)\textsuperscript{-1}, as shown in Fig. \ref{fig:solo_data} and listed in Col. 7 of the table. Most of the SEE events measured by Solar Orbiter are related to small jets and brightening, as observed with EUV imaging, related to B- and C- class flares, based on GOES X-ray 1-8 {\AA} channel, as indicated in Col. 4. Three events (5, 9 and 12) presented peak intensities above $\sim$10\textsuperscript{3} (cm\textsuperscript{2} sr s MeV)\textsuperscript{-1}. The event \#9 is related to a gradual and long duration increase in the X-ray flux given by GOES and to a moderate CME speed. It also presents a good nominal connection to the source region (CA $\sim$ -15$^{\circ}$). The two most intense events of the period (\#5 and \#12) are associated with fast CMEs and CME-driven shocks. The electron intensities are still moderate in both events probably because the nominal magnetic connection between Solar Orbiter and the source region was poor (|CA| $\gtrsim$75$^{\circ}$). For these two events, the flare location was estimated based on the longitude and latitude of the CME apex given by the 3D reconstruction (not shown here), as the respective solar sources were not visible from Earth or STEREO-A point of view. This is indicated with (a) in Col. 4 of Table \ref{Table:SEP_list_Solo}. These two SEP events, 2022 February 15 (event \#5) and 2022 March 21 (event \#12), are widespread events (indicated with * in Col. 1), and were observed by several spacecraft widely separated in heliolongitude and at different radial distances from the Sun. In total, three events in the list can be classified as widespread events, but, as explained in Sect. \ref{sec:MESSENGER SEE list}, the number of widespread events could be larger.

\begin{figure*}
\centering
\includegraphics[scale=0.6] {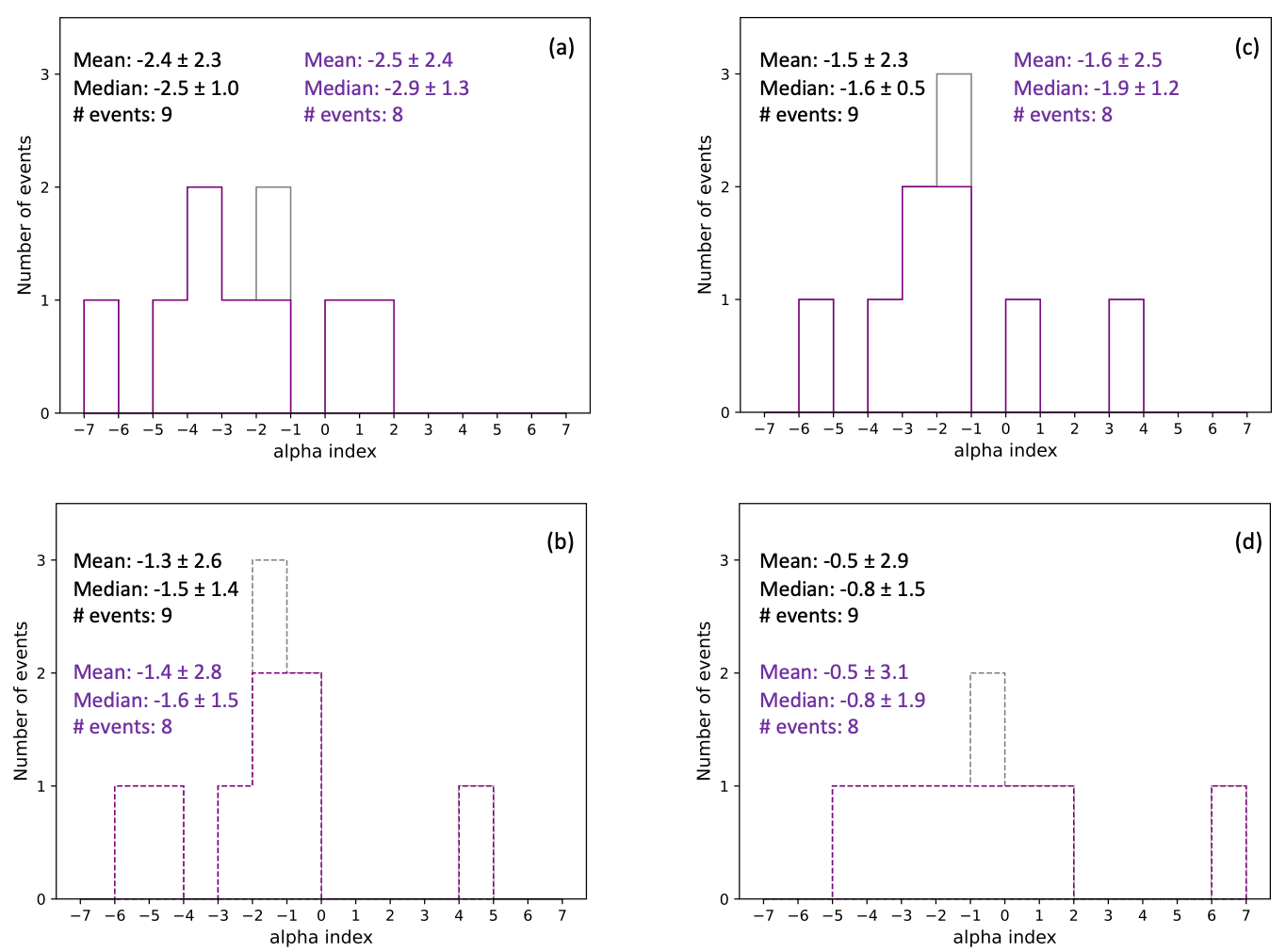}
     \caption{Histograms of the  indices $\alpha$ for a radial dependence $\propto$ R\textsuperscript{$\alpha$} of the quasi-relativistic electron intensities in SEE events measured by Solar Orbiter. \textit{(a)} Gray (purple) color indicates the $\alpha$ indices for those SEP events for which the longitudinal separation of the nominal footpoints of Solar Orbiter and the respective spacecraft observing near 1 au were $\leq$ 35$^{\circ}$ (< 20$^{\circ}$). \textit{(b)} Same as in (a) but including a correction for the different connection angle. \textit{(c),(d)} The same as in (a) and (b) but including a calibration factor of 1.3 on STEREO measurements. The legend shows both the mean and standard deviation and the median and median absolute deviation. Details given in the text.      }
     \label{fig:alpha_histograms_solo}
\end{figure*}

\subsection{Radial dependence of the peak intensity in SEE events measured by Solar Orbiter}
\label{sec:SolO_radial_dependence}

In this section we analyse the radial dependence of the electron peak intensity of the SEE events listed in Table \ref{Table:SEP_list_Solo}. During February 2022, Solar Orbiter was consistently in nominal magnetic connection with STEREO-A (left panel of Fig. \ref{fig:solar_mach}). During March 2022, this good nominal magnetic connectivity slowly shifted from STEREO-A at the beginning of the month to ACE and Wind spacecraft at the end of the period of study (right panel of Fig. \ref{fig:solar_mach}). Thus, in 11 of the 12 SEE events listed here the longitudinal separation of the nominal footpoints of Solar Orbiter and the respective observations near 1 au is < 20$^{\circ}$, as indicated with ++ in Col. 1 of Table \ref{Table:SEP_list_Solo}. The specific connection angle difference between the respective spacecraft near 1 au and Solar Orbiter is given in parenthesis in Col. 11 of the list. 

For near 1 au observations, we used data from STEREO-A/SEPT and ACE/EPAM/DE as already explained in Sect. \ref{sec:Radial_depence_data_source}. After the solar superior conjunction of the STEREO spacecraft (from January to August 2015), STEREO-A spacecraft was rolled 180 degrees about the spacecraft-Sun line in order to allow the high gain antenna to remain pointing at Earth. Consequently, the nominal pointing directions of the SEP suite of instruments are now different from originally intended and therefore we used omnidirectional averaging to measure the peak intensities. The electron peak intensities are listed in Col. 10 of Table \ref{Table:SEP_list_Solo}. The units for both the peak intensity and the pre-event background intensity (in parenthesis) are particles (cm\textsuperscript{2} sr s MeV)\textsuperscript{-1}. As indicated with "several events mixed" in Col. 10 for events number 5, 9, and 10, there was no clear association between the increase in electron intensities measured by STEREO-A and the solar event indicated in Cols. 1-3. Thus, we used the nine remaining events for the radial dependence analysis.  

As we did for MESSENGER SEE events, we directly compared EPD/EPT on board Solar Orbiter with STEREO/SEPT data without using any scaling correcting factor, but we divided ACE data by the inter-spacecraft calibration factor of 1.3 with STEREO, as discussed in Sect. \ref{sec:Radial_dependence}. This is noted in Col. 10 of Table \ref{Table:SEP_list_Solo} with (\&). However, a preliminary comparison of electron measurements by ACE/EPAM/DE and EPT on board Solar Orbiter, shows that both instruments measured very similar intensities at energies near $\sim$70-100 keV \citep{Gomez-Herrero2022}. This comparison was made during the SEE events that occurred when Solar Orbiter was relatively close to Earth near the  Earth flyby on 27 November 2021. Therefore, we also used a multiplying factor of 1.3 on STEREO data to estimate the radial dependence based on the inter-calibration factor discussed in Sect. \ref{sec:Radial_dependence}, using ACE data with no factor. We also applied the correction for the small longitudinal effect present in the sample, as discussed in Sect. \ref{sec:Radial_dependence}.

Figure \ref{fig:solo_radial_depend} shows the peak intensity as a function of heliocentric distance for Solar Orbiter SEE events. The gray (purple) lines connect the peak intensities in the prompt component of the events, where the separation of the nominal footpoints of Solar Orbiter and the respective spacecraft observing near 1 au were < 35$^{\circ}$ (< 20$^{\circ}$). The index $\alpha$ for each SEE event is shown in the legend, along with the corrections due to the longitudinal effect. We also found an event-to-event variability, with $\alpha$ values varying between $\sim$-6 (event \#4) and $\sim$2 (event \#2). The difference between Figs. \ref{fig:solo_radial_depend}a and \ref{fig:solo_radial_depend}b is that Fig. \ref{fig:solo_radial_depend}a compares directly Solar Orbiter data with STEREO-A, dividing ACE data by 1.3, and Fig. \ref{fig:solo_radial_depend}b directly compares Solar Orbiter and ACE data, multiplying STEREO-A data by 1.3. 

Figure \ref{fig:alpha_histograms_solo} shows the histogram of the alpha indices for the radial dependence $\propto$ R\textsuperscript{$\alpha$} presented in Fig. \ref{fig:solo_radial_depend}. Figure \ref{fig:alpha_histograms_solo}a (b) shows the result based on no inter-spacecraft calibration factor between Solar Orbiter and near-Earth spacecraft for the peak (Corrected) intensity. Figure \ref{fig:alpha_histograms_solo}c (d) shows the result based on multiplying STEREO-A data by an intercalibration factor of 1.3 for the peak (Corrected) intensity. 
The legend in Fig. \ref{fig:alpha_histograms_solo}a shows that the median and the MAD of the $\alpha$ index, if a radial dependence R\textsuperscript{$\alpha$} is assumed, are $\alpha$\textsubscript{Med}=-2.5$\pm$1.0, based on 9 SEE events and not including the corrections due to the intercalibration factor or the longitudinal effect. The correction in the median $\alpha$ index due to the intercalibration factor and longitudinal effect is $\sim$36\% each ($\alpha$\textsubscript{Med}=-1.6$\pm$0.5 and $\alpha$\textsubscript{Med}=-1.5$\pm$1.4, respectively). When considering both corrections, the median of the $\alpha$ index is $\alpha$\textsubscript{Med}=-0.8$\pm$1.9. A summary of the $\alpha$ indices for the different corrections and subsamples is presented in Cols. (5)-(6) of Table \ref{table:summary of alpha indices}.




\begin{figure*}
\centering
  \resizebox{1.0\hsize}{!}{\includegraphics{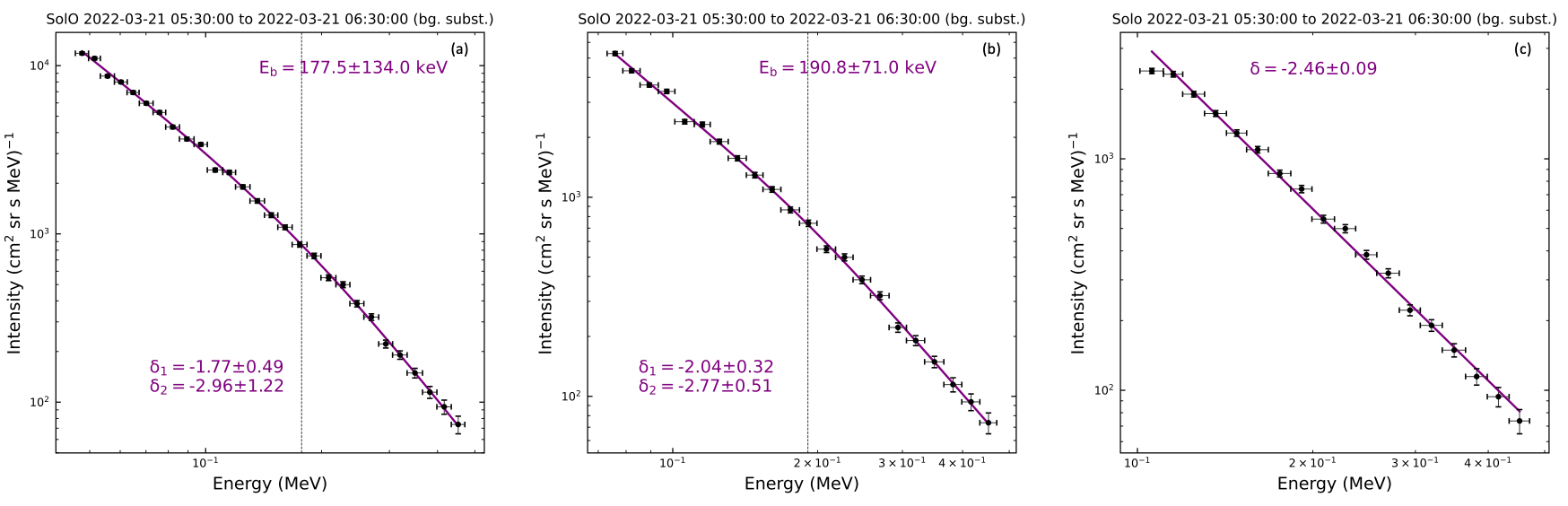}}
     \caption{Example of Solar Orbiter solar electron peak spectra. \textit{(a)} Solar Orbiter omnidirectional spectrum showing a broken power-law for the fitting. \textit{(b) and (c)} Same spectrum as in (a) but respectively using only energies above $\sim$70 keV and $\sim$100 keV for the fitting, showing a double and single power-law, respectively. }
     \label{fig:Solo_spectra}
\end{figure*}
\begin{figure}
\centering
  \resizebox{\hsize}{!}{\includegraphics{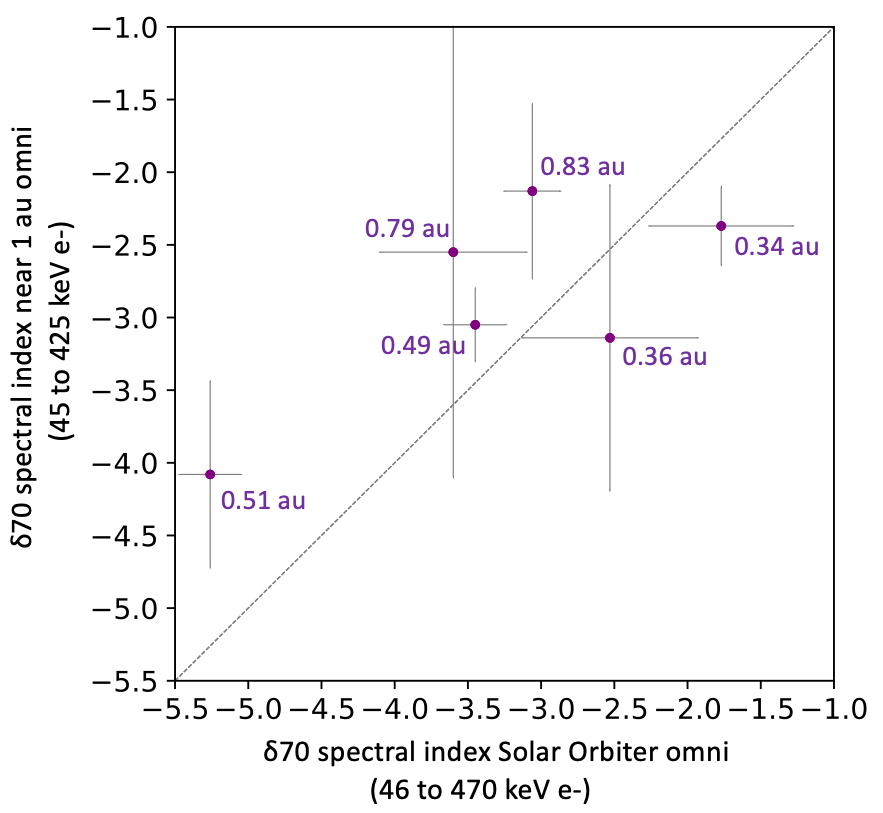}}
     \caption{$\delta$70 spectral indices  near 1 au against the spectral indices at Solar Orbiter. The legend shows the radial distance of Solar Orbiter for each event. }
     \label{fig:Solo_spectra_dif}
\end{figure}

\subsection{Peak-intensity energy spectra and their radial dependence in SEE events measured by Solar Orbiter}
\label{sec:SolO peak spectra}
In this section we present the selection of SEE events measured by Solar Orbiter where the peak-intensity energy spectra could be determined. We also study the radial dependence of the spectral indices. As presented in Sect. \ref{sec:Solo_data_source}, the EPT instrument on board Solar Orbiter measures electrons from $\sim$30 keV to $\sim$470 keV divided into 34 energy bins, in four different pointing directions (Sun, anti-Sun, north and south). We used the omnidirectional averaging to compare with STEREO-A data, as discussed in Sect. \ref{sec:SolO_radial_dependence}. The EPT energy channels below $\sim$45 keV are affected by an instrumental
effect which we are currently investigating. It leads to
substantial variations in the count rates of these energy channels and
affects the electron spectra. Because we are particularly interested in the
characterization of the spectral transitions and because these variations are likely to influence the spectral fit, we excluded these low-energy bins from our analysis. Then, the energies used in the spectrum analysis are $\sim$46 keV to $\sim$470 keV, divided into 28 bins. To compare with MESSENGER results, we followed the same procedure as explained in Sect. \ref{sec:MESSENGER_peak_spectra} regarding the hourly-average TOM spectra, in this case based on the 73-111 keV energy range. We also subtracted the pre-event background level and checked the significance level, as defined in Sect. \ref{sec:MESSENGER_peak_spectra}, and the ion contamination. From the original list of 12 events, we discarded three events following the criteria explained in Sect. \ref{sec:MESSENGER_peak_spectra}. 

For the remaining nine events, we followed the fitting procedure as explained by \cite{Strauss2020}. The $\delta$ spectral indices found in the energy range around 70 keV (200 keV), namely $\delta$70 ($\delta$200), are given in Col. 8 (9) of Table \ref{Table:SEP_list_Solo}. Only four events (number 5, 9, 10 and 12) showed electron intensity increase above 200 keV. Four events (number 5, 9, 11 and 12) could be fitted using a broken power-law, where the mean spectral transition is 111$\pm$53 keV.  
Figure \ref{fig:Solo_spectra}a shows the omnidirectional peak-intensity energy spectrum fitting for event \#12 (2022/03/21), when Solar Orbiter's radial distance from the Sun was 0.34 au, showing a double power-law. The spectral index below ($\delta$\textsubscript{1}) and above ($\delta$\textsubscript{2}) the spectral transition E\textsubscript{b} (vertical dashed line) are shown in the legend. The uncertainties are calculated for a confidence interval of 95\%. Figure \ref{fig:Solo_spectra}b (\ref{fig:Solo_spectra}c) shows the spectral fitting for the same event (2022/03/21), but using only the energy bins above $\sim$70 ($\sim$100) keV, which is a similar starting energy for the fittings done with MESSENGER data. We note that in the case of the MESSENGER spectra, there is only one bin (channel) covering $\sim$70-100 keV energies. The fitting still resembles a double power-law for Fig. \ref{fig:Solo_spectra}b but not for Fig. \ref{fig:Solo_spectra}c. 


The mean of the <$\delta$70> indices for nine events measured by Solar Orbiter is -2.9$\pm$0.3, while the mean of the <$\delta$200> for the four events showing >200 keV electron intensity enhancements is -2.7$\pm$0.5. We note that the mean of the <$\delta$70> indices for the four events that show increases above 200 keV is -1.98$\pm$0.36, harder than the <$\delta$200> of -2.7$\pm$0.5, as found in previous studies \cite{Dresing2020}. To analyse the possible effect of the IP scattering on the energy spectrum, we compared the Solar Orbiter spectra with near 1 au measurements, as described in Sect. \ref{sec:spectra_comparison}, using the $\delta$70 index. From the nine events measured by Solar Orbiter where the spectra could be determined, only six events could be used for the radial dependence study. As discussed in Sect. \ref{sec:SolO_radial_dependence}, in events number 5, 9, and 10 no clear association between the increase in electron intensities measured by STEREO-A and the solar event was found.

Figure \ref{fig:Solo_spectra_dif} shows the $\delta70$ spectral indices measured near 1 au against the spectral indices at Solar Orbiter. Although we have only six SEE events, it is interesting to note that the position of the points with respect to the dashed line (representing equal indices) depends on the distance from the Sun of the Solar Orbiter spacecraft. The two events below the line correspond to events \#11 (0.36 au) and \#12 (0.34 au), where the spectral indices near 1 au are softer than near 0.3 au. 

\section{Summary and discussion}
\label{sec:summary_discussion}
We presented a list of 61 SEE events measured by the MESSENGER mission from 2010 to 2015 when the heliocentric distance of the spacecraft varied from 0.31 au to 0.47 au. As a preamble for future studies using data from new missions exploring the innermost regions of the heliosphere, we also included in the study a reduced list of 12 SEE events measured by Solar Orbiter when the spacecraft was close to its first perihelion passage, in particular when the Solar Orbiter's radial distance varied from 0.34 to 0.83 au. For each of the SEE events, we identified the respective solar origin (flare location, type III radio burst onset), estimated the nominal connection angle to the solar source, measured the peak intensity, and calculated the peak-intensity energy spectrum when possible. 

Due to the elevated background level of the MESSENGER/EPS instrument, the majority of the SEE events measured by MESSENGER presented high peak intensity levels, with $\sim$1 MeV electron intensity enhancements (37 events), and being widespread in heliolongitude (32 events). For most of these events (56 out of 61), a CME-driven shock was detected in EUV and white-light coronagraph images. There were four events showing peak intensities above 10\textsuperscript{7} (cm\textsuperscript{2} sr s MeV)\textsuperscript{-1}. Among them, the SEP events on 2011 June 4 and 2012 March 7, discussed in detail by \cite{2013Lario_intense}, and the widespread event on 2013 August 19, studied in detail by \cite{2021Rodriguez-Garcia}. In the case of the SEE events measured by Solar Orbiter, most of them (8 out 12) were related to small jets and brightening presenting peak intensities below 10\textsuperscript{3} (cm\textsuperscript{2} sr s MeV)\textsuperscript{-1}. The two most intense events of the period, on 2022 February 15 and March 21, were associated with fast CMEs and CME-driven shocks. 
 


We derived the radial dependencies of the near-relativistic electron peak intensities in 28 SEE events measured by MESSENGER/EPS and in 9 SEE events measured by Solar Orbiter/EPT, when a spacecraft near 1 au was closely aligned with them along the nominal Parker spiral. In the case of MESSENGER, we find that the radial dependence can be represented as $\propto$ R\textsuperscript{$\alpha$}, where the median alpha index is $\alpha$\textsubscript{Med}=-3.3$\pm$1.4, including both the corrections by the small longitudinal distance between the footpoints and the intercalibration factor between the spacecraft. However, their effect on the median $\alpha$ index is small, namely 12\% and 8\%, respectively. In order to estimate how the anti-sunward MESSENGER/EPS field of view limitation affects our results, we compared the ratio between intensities of the sunward and anti-sunward looking instruments obtained with telescopes on other spacecraft. The median ratio for the SEE events analysed in the present work is I\textsubscript{max\_sun}/I\textsubscript{max\_asun} =1.1±0.2 for STEREO, and =1.3±0.5 for Solar Orbiter. We therefore expect that the restricted field of view of  MESSENGER has an effect on the median indices given in this study that lies within the given uncertainties. 

In the case of Solar Orbiter, the median $\alpha$ index is $\alpha$\textsubscript{Med}=-0.8$\pm$ 1.5. We note the reduced  number (9) of events  included in the sample and the presence of an event with a positive slope. Due to the smaller radial separation between Solar Orbiter and STEREO-A for the majority of the events in comparison with the MESSENGER study, both the longitudinal effect and the intercalibration factor play a stronger effect (36\% each) on the radial dependence analysed here than in the MESSENGER study. Although the $\alpha$ indices derived from MESSENGER and Solar Orbiter data are comparable within the uncertainties, the differences might also result from the type of events included in the two samples. We mainly measured large 
intense events with MESSENGER while the event sample for Solar Orbiter mostly (8 out of 9) includes small impulsive events (peak intensities $\sim$10\textsuperscript{2} (cm\textsuperscript{2} sr s MeV)\textsuperscript{-1}). This selection may have an impact in the derived indices that cannot be evaluated in this work. Future studies covering different phases of the solar activity and with larger radial separations between the spacecraft could investigate further the radial dependences for different kind of samples.

We note that there is an event-to-event variability in the radial dependence of the peak intensities, with $\alpha$ values varying between $\sim$-9 and $\sim$0 in the case of MESSENGER and between $\sim$-6 and $\sim$2 in the case of Solar Orbiter. A steeper or flatter radial dependence than R\textsuperscript{-3}, might be due to pre-existent transient solar wind structures, such as IP shocks and/or ICMEs and stream interaction regions (SIRs), and/or the variability of the scattering processes undergone by the particles during their transport in IP space \citep{2007Lario}. We note that the presence of these intervening structures, apart from having an effect on the transport of energetic particles, may also modify the estimated CA. Examples of these scenarios for the MESSENGER SEE events studied here can be found in published studies, such as \cite{2013Lario_intense} for events \#5 (2011/06/04b) and \#19 (2012/03/07); \cite{Lario2013} for SEE event \#2 (2010/08/18), \#4 (2011/06/04a), and \#13 (2011/11/17); and \cite{2021Rodriguez-Garcia} for event \#36 (2013/08/19). 
The actual heliospheric magnetic field configuration in which energetic electrons propagate varies from event to event. This may affect the radial evolution of electron intensities, but can only be assessed in individual event studies. The statistical approach of our work describes an average behaviour.



We also analysed the peak-intensity energy spectra in 42 SEE events measured by MESSENGER/EPS, where the spectra could be determined. The solar origin of most of these events (40 out of 42) involved a CME-driven shock seen by EUV and white-light coronagraph images. For all the events, the peak intensity energy spectra resembled a single power-law. In some of the Solar Orbiter events, we could fit the spectra using a broken power-law, including two events when the spacecraft was near 0.3 au. We also fitted one of these events, namely the event observed on 2022 March 21, using only the energy bins above $\sim$70 keV and $\sim$100 keV. The fittings show a broken and single power-law, respectively. The difference between the fittings is mainly the number of bins and the range of energies included. Thus, the identification of a transition in the spectra is strongly conditioned by the experimental data set, particularly by the spectral resolution (i.e. number of energy bins) and the extension of the energy interval covered by the instrument. Thus, the lack of bins available at energies below $\sim$70 keV for MESSENGER data and the width of the energy bins above this energy, could be the main reason behind not finding a spectral transition.

The mean and median spectral index of the 42 SEE events measured by MESSENGER are <$\delta$>=$\delta$\textsubscript{Med}=-1.9$\pm$0.3. 
For comparison, we considered two previous studies that analysed the forward spectra of  near-relativistic solar energetic electron events near 1 au.  \cite{2009Krucker} studied 62 impulsive electron events within energies from 1 to 300 keV measured by the Wind spacecraft, finding a <$\delta$200> mean value of -3.6$\pm$0.7. \cite{Dresing2020} analysed 781 near-relativistic solar energetic electron events measured by both STEREO spacecraft. They find a <$\delta$200> mean value of -3.5$\pm$1.4.  Thus, the MESSENGER backward spectra is much harder than the forward spectra analysed by \cite{2009Krucker} and \cite{Dresing2020}. With regard to estimate how the anti-sunward MESSENGER/EPS field of view limitation  affects our spectra results, we compare the ratio between spectral indices of the sunward and anti-sunward looking instruments obtained with telescopes on other spacecraft. The  ratio for the SEE events analysed in the present work is $\delta$200\textsubscript{forward}/ $\delta$200\textsubscript{backward}=1.1$\pm$0.1 for STEREO, and =1.10$\pm$0.02 for a reduced sample of three events measured by Solar Orbiter. We therefore expect that the hypothetical forward spectra of MESSENGER could also be $\sim$10\% softer than the backward spectra, leading to a mean spectral value of <$\delta$>\textsubscript{forward}=-2.1$\pm$0.3. This estimated mean value is still harder than that using near 1 au data mentioned above.

The difference between the mean of the spectral indices measured in this study and in \cite{2009Krucker} or \cite{Dresing2020} might be related to the characteristics of the events measured by MESSENGER. Because of the elevated background intensity level of the EPS instrument, the majority of events are very intense. Additionally, the MESSENGER events showed wide angular particle spreads (32 out of 42), were accompanied by coronal CME-driven shocks (40 out of 42), and extended to high energies exhibiting $\sim$1 MeV electrons intensity enhancements (37 out of 42). \cite{2022Dresing} analysed 33 electron energetic events that were related to the presence of coronal pressure waves. They derive a mean spectral index of <$\delta$200> = -2.5$\pm$0.3. Thus, the subsample of events with coronal pressure waves contains those events with the hardest energy spectra in the whole sample of electrons events observed by STEREO during solar cycle 24. However, MESSENGER spectra are still harder than this subsample measured near 1 au, even when only using the higher energy bins of MESSENGER/EPS for the spectral fitting. However the reliability of this comparison is low due to the very large uncertainties on the fitting using only three wide bins, as discussed in Sect. \ref{sec:MESSENGER_peak_spectra}. 

The peak-intensity energy spectra analysis using Solar Orbiter data is in agreement with the results presented above for MESSENGER regarding the softening of the spectra between 0.3 au and 1 au.
The mean spectral indices at both energies, $\delta$70 (9 events, <$\delta$70>=-2.9$\pm$0.3) and $\delta$200 (4 events, <$\delta$200>=-2.7$\pm$0.5) measured at Solar Orbiter are also harder than those reported by \cite{Dresing2020} using near 1 au measurements, but softer than MESSENGER results. If the scattering conditions for SEEs are energy dependent, namely the higher the energy of the electrons, the more frequent the scattering processes undergone by the particles \citep{2003Droge,Strauss2020}, this can result in a significant change of the energy spectrum during the transport of the particles from the Sun to the observer.
We note that Solar Orbiter distance from the Sun varied from 0.34 to 0.83 au. Thus, the effect of IP scattering processes, when present, might affect the Solar Orbiter sample in a higher degree than for the MESSENGER sample (always closer than 0.47 au), softening the spectra. Interestingly, the softening of the $\delta$70 spectra index might depend on the distance of the spacecraft closer to the Sun that is used to compare with near 1 au measurements. For measurements when Solar Orbiter was near 0.3 au, the spectral index near 1 au is softer than the spectral index measured at Solar Orbiter, as we found for MESSENGER events. This distance dependence might be further explored in the future, as we progress into solar cycle 25.

We also further investigated the radial dependence of the peak-intensity energy spectra, comparing MESSENGER spectral index with near 1 au forward (backward) spectra for a subsample of 20 (18) SEE events. We found that the backward spectra near 0.3 au is harder than the forward (backward) spectra near 1 au by a median factor of $\sim$20\% ($\sim$10\%). Moreover, considering the ratio between forward and backward spectra near 0.3 au, as discussed above, the forward near 1 au spectra might be 10\% softer than the forward spectra near 0.3 au. However, there is no intercalibration among the different energies measured by MESSENGER/EPS, STEREO/SEPT and/or Wind/3DP instruments, which would allow us to discard any systematic effects that could influence their comparison.  


This statistical study of intense SEE events at heliocentric distances near 0.3 au is relevant and timely. The radial dependences of SEE intensities are used in Space Weather to determine the particle radiation environment at different distances from the Sun, and they can also be used to constrain the models applied to understand the different scenarios of acceleration and transport of SEE in the heliosphere. The radial dependence of the electron energy spectra and the presence of potential spectral transitions might be used in (and explained with) the different modelling efforts at this respect. The analysis and outcomes presented here might be further investigated with data from the new ongoing missions exploring the innermost regions of the heliosphere, such as Parker Solar Probe, Solar Orbiter and BepiColombo, together with data from near 1 au spacecraft. By using these new multi-spacecraft observations and as we progress into solar cycle 25, we could measure more intense events and increase the statistics, which will allow a reduction of the uncertainties. We could also analyse those events departing from the inferred radial dependence of peak intensities and energy spectra. Moreover, future cross calibration studies during close approaches of the different spacecraft will grant a more accurate inter-comparison of the data sets from different instruments.


\section{Conclusions}
\label{sec:Conclusions}
In this paper we studied the radial dependence of the electron peak intensity and peak-intensity energy spectrum for SEE events measured by the MESSENGER mission from 2010 to 2015, when MESSENGER's heliocentric distance varied between 0.31 and 0.47 au. We also analysed a reduced list of SEE events measured by Solar Orbiter during February and March 2022, when Solar Orbiter's heliocentric distance varied from 0.34 to 0.83 au. The three main conclusions derived from this study are as follows: 

\begin{enumerate}

 \item[1.] Most of the selected events measured by MESSENGER/EPS are very intense, accompanied by a CME-driven shock, extended to high ($\sim$1 MeV) energies and are widespread in longitude. The sample is biased towards large intense SEE events, because of the high background level of this instrument.\newline
 \item[] The two main conclusions derived from the analysis of the large SEE events measured by MESSENGER, which  are generally supported by Solar Orbiter's data results, are:
 \end{enumerate}

 \begin{enumerate}
 \item [2.] There is a wide variability in the radial dependence of the electron peak intensity between $\sim$0.3 au and $\sim$1 au, but the peak intensities of the energetic electrons decrease with radial distance from the Sun in 27 out of 28 events. On average and within the uncertainties, we find a radial dependence consistent with R\textsuperscript{-3}. 
 
 \item [3.] The electron spectral index found in the energy range around 200 keV ($\delta$200) of the backward-scattered population near 0.3 au is harder in 19 out of 20 (15 out of 18) events by a median factor of $\sim$20\% ($\sim$10\%) when comparing to the anti-sunward propagating beam (backward-scattered population) near 1 au.
 
\end{enumerate}

\begin{acknowledgements} The UAH team acknowledges the financial support by the Spanish Ministerio de Ciencia, Innovación y Universidades FEDER/MCIU/AEI Projects ESP2017-88436-R and PID2019-104863RB-I00/AEI/10.13039/501100011033 and by the European Union’s Horizon 2020 research and innovation program under grant agreement No. 101004159 (SERPENTINE). LRG is also supported by the European Space Agency, under the ESA/NPI program and thanks Nariaki Nitta for his help and Karl Ludwig Klein for the helpful comments and discussion that contributed to improve the quality of the manuscript. DL acknowledges support from NASA Living With a Star (LWS) programs NNH17ZDA001N-LWS and NNH19ZDA001N-LWS, and the Goddard Space Flight Center Internal Scientist Funding Model (competitive work package) program and the Heliophysics Innovation Fund (HIF) program. LAB acknowledges the support from the NASA program NNH17ZDA001N-LWS (Awards Nr. 80NSSC19K0069 and 80NSSC19K1235). AK acknowledges financial support from NASA NNN06AA01C (SO-SIS Phase-E) contract. ND is grateful for support by the Turku Collegium for Science, Medicine and Technology of the University of Turku, Finland. Solar Orbiter is a mission of international cooperation between ESA and NASA, operated by ESA. The authors acknowledge the different Wind, ACE, SOHO, STEREO, MESSENGER, Solar Orbiter instrument teams, and the STEREO and ACE science centers for providing the data used in this paper. The MESSENGER data used in this study is available online. The EPD/EPT sensor was built with support from the German Space Agency,
DLR, under grants 50OT0901, 50OT1202, 50OT1702, and 50OT2002. The radio spectrograms used in this study are provided by the Observatoire de Paris-Meudon (\url{http://secchirh.obspm.fr/}). 

\end{acknowledgements}

\begin{flushleft}

\textbf{ORCID iDs} 
\vspace{2mm}

Laura Rodríguez-García \orcid{https://orcid.org/0000-0003-2361-5510}

Raúl Gómez-Herrero \orcid{https://orcid.org/0000-0002-5705-9236}

Nina Dresing \orcid{https://orcid.org/0000-0003-3903-4649}

David Lario
\orcid{https://orcid.org/0000-0002-3176-8704}

Ioannis Zouganelis
\orcid{https://orcid.org/0000-0003-2672-9249}

Laura Balmaceda \orcid{https://orcid.org/0000-0003-1162-5498}

Athanasios Kouloumvakos \orcid{https://orcid.org/0000-0001-6589-4509}

Annamaria Fedeli \orcid{https://orcid.org/0000-0001-9449-4782}

Francisco Espinosa Lara \orcid{https://orcid.org/0000-0001-9039-8822}

Ignacio Cernuda
\orcid{https://orcid.org/0000-0001-8432-5379}

George Ho
\orcid{https://orcid.org/0000-0003-1093-2066}

Robert F. Wimmer-Schweingruber \orcid{https://orcid.org/0000-0002-7388-173X}

Javier Rodríguez-Pacheco
\orcid{https://orcid.org/0000-0002-4240-1115}

\end{flushleft}
%
%
\bibliographystyle{bibtex/aa}

\end{document}